\documentclass{aastex}
\usepackage{emulateapj5}
\def\nad{\mbox {Na~I}}
\def\i{\mbox {\rm IRAS}}
\def\h1{\mbox {\rm HI}}
\def\x{\mbox {\rm X-ray~}}
\def\X{\mbox {\rm X-Ray~}}
\def\fig{{Figure}}

\def\deg{\mbox {$^{\circ}$}~}
\def\msun{\mbox {${\rm ~M_\odot}$}}

\def\lsun{\mbox {${~\rm L_\odot}$}}

\def\msunyr{\mbox {$~{\rm M_\odot}$~yr$^{-1}$}}
\def\msunyrk2{\mbox {$~{\rm M_\odot}$~yr$^{-1}$~kpc$^{-2}$}}
\def\msunpc2{\mbox {${\rm ~M_\odot ~pc}^{-2}$}}
\def\angs{\mbox {~\AA}}
\def\lya{\mbox {Ly$\alpha$~}}
\def\Ha{\mbox {H$\alpha$~}}

\def\line{\mbox {~$\lambda$}}

\def\n{NGC~}
\def\asec{\ifmmode {'' }\else $''~$\fi}  % arc sec
\def\amin{\ifmmode {' }\else $'~$\fi}    % arc min
\def\sles{\lower2pt\hbox{$\buildrel {\scriptstyle <}
   \over {\scriptstyle\sim}$}} % approximately less than
\def\sgreat{\lower2pt\hbox{$\buildrel {\scriptstyle >}
    \over {\scriptstyle\sim}$}} % approximately greater than
\def\col{\mbox {\rm ~cm$^{-2}$} }
\def\kms{\mbox {~km~s$^{-1}$} }

\def\cm3{~cm$^{-3}$}
\def\cm5{~cm$^{-5}$}

\def\um{\mbox {$\mu${\rm m}} }

\def\et{{\rm et\thinspace al.}\ }   % et al.
\def\apj{ApJ}
\def\apjs{ApJS}

\def\aj{AJ}
\def\mn{MNRAS}

\def\aa{A\&A}

\begin{document}

%\title{Galactic Superwinds in Ultraluminous Infrared Galaxies:
%NaD Absorption Line Profiles in Keck II ESI Spectra}

\title{Mapping Large-Scale Gaseous Outflows in Ultraluminous
Galaxies with Keck~II ESI Spectra: Variations in Outflow Velocity with
Galactic Mass\altaffilmark{1}}

\author{Crystal L. Martin\altaffilmark{2,} \altaffilmark{3}}
\affil{University of California, Santa Barbara}
\affil{Department of Physics}
\affil{Santa Barbara, CA, 93106}
\email{cmartin@physics.ucsb.edu}

\altaffiltext{1}{Data presented herein were obtained at the W.M. Keck
Observatory, which is operated as a scientific partnership among the 
California Institute of Technology, the University of California and the
National Aeronautics and Space Administration.  The Observatory was made
possible by the generous financial support of the W.M. Keck Foundation.
}
\altaffiltext{2}{Packard Fellow}
\altaffiltext{3}{Alfred P. Sloan Research Fellow}

% {Submitted Version 2004 Aug 12}
% 8/27/04: Revised to reflect comments from Evan, Carl, and Eliot 

\begin{abstract} 

Measurements of interstellar \nad\ $\lambda \lambda 5890, 96$ absorption lines
in 18 ultraluminous infrared galaxies (ULIGs) have been combined with
published \nad\ data, in order to reassess the dependence of galactic
outflow speeds on starburst luminosity and galactic mass.  The Doppler
shifts reveal outflows of relatively cool gas in 15 of 18 ULIGs with 
an average outflow speed at the line center of  $330 \pm 100$\kms.
The relation between outflow speed and star formation rate (SFR), defined
by the distribution's upper envelope over four orders of magnitude
in SFR, demonstrates that winds from more luminous starbursts accelerate 
interstellar gas to higher speeds as  roughly $v \propto SFR^{0.35}$.
This result is surprising since, in the traditional model for starburst-driven
winds, these relatively cool gas clouds are accelerated by the ram pressure of
a hot, supernova-heated wind that exhibits weak (if any) \x temperature
variation with increasing galactic mass. The lack of evidence for 
much hotter winds is partly a sensitivity issue; but the  \nad\
velocities in ultraluminous starbursts actually are consistent with 
acceleration by the tepid wind indicating a hotter component is unlikely to 
dominate the momentum flux. The \nad\ 
velocities in the dwarf starburst winds do not reach the terminal velocity of 
a hot wind at the measured temperature of $kT \sim 0.73$~keV  -- 
a result which  
could be interpreted simply as evidence that the hot superbubbles are too 
confined in dwarf starbursts to generate a free-flowing wind. A 
dynamically-motivated scenario, however, is that the dwarf starburst winds 
simply lack enough momentum to accelerate the clouds to the velocity of the hot
wind. Among the subsample of starbursts with well-constrained dynamical 
masses, the terminal outflow velocities are found to always approach
the galactic escape velocity. Motivated by a similar scaling relation
for {\it stellar} winds, the galactic Eddington luminosity for dusty 
starbursts is shown to be within the range measured for ULIGs. If
radiation pressure on dust grains, coupled to the cool wind, is indeed
important for galactic wind dynamics,  then feedback will be stronger
in massive galaxies than previously thought helping shape the high-mass 
end of the galaxy luminosity 
function. Regardless of the nature of the acceleration mechanism in ULIGs, the 
mass flux of cool gas estimated from these data demonstrates that 
starburst-driven winds transport significant gas during the assembly stage 
of field elliptical galaxies -- a factor which helps explain the
rapid decline in SFR in these systems inferred from elemental abundance ratios.
\end{abstract}

%The terminal wind velocities, typically  $\sim 750$\kms\ in the ULIGs, 
%scale nearly linearly with the galactic circular velocity and approach
%the halo escape velocity.

\subjectheadings{
galaxies: formation --- 
galaxies: evolution ---
galaxies: fundamental parameters ---  
ISM: kinematics and dynamics
ISM: structure
ISM: evolution
}


\section{Introduction}

Galactic-scale gaseous outflows  are ubiquitous in starburst galaxies
at all cosmic epochs (Heckman, Armus, \& Miley 1990; Pettini \et 2001;
Shapley \et 2003).
Powered by supernova explosions, perhaps with additional help from AGN in 
very luminous galaxies, these galactic winds transport  heavy elements, 
interstellar dust, and energy into galaxy halos and the intergalactic 
medium.  The significance for galaxy evolution depends on details of the
coupling between the available energy, both radiative and mechanical,
and the interstellar gas.  Observations, which describe this interplay 
empirically, illuminate the dominant physical processes and provide 
quantitative constraints for galaxy evolution models.

This paper directly addresses feedback at the high mass end of the 
galaxy luminosity function via measurements of outflow speeds in
ultraluminous starbursts. These new observations of Na I and K I  
(ionization potential of 5.1~eV and 4.3~eV respectively)
interstellar absorption reveal relatively cold gas clouds advected
into the outflow at the shear interface between the hot wind and
the quiescent gas disk. Material evaporated from these clouds could
be the source of the mass-loading required to explain the large 
X-ray surface brightnesses of galactic winds (e.g. Suchkov \et 1994; 
Suchkov \et 1996; Strickland \et 2002; Martin, Kobulnicky, \& Heckman 2002). Unlike 
emission lines, absorption lines leave no ambiguity between infall and outflow 
and are sensitive to extended, low density gas. The results reported here 
complete a larger effort to describe the low-ionization absorption 
kinematics over four orders of magnitude in star formation rate
(Heckman \et 2000, luminous infrared galaxies (LIGs);  Schwartz \& Martin 
2004, dwarf starbursts). The observations have little overlap with the
Rupke \et (2002) sample of ULIGs due to different selection criteria (see \S 2).

Feedback is expected to have a large impact on dwarf galaxies owing to their 
shallow gravitational potential.  It can explain the absence of young galaxies 
with very low mass, $V_{rot} \sles\ 10 - 20$\kms, 
after reionization (Barkana \& 
Loeb 1999), the slope of the galaxy luminosity function (Dekel \& Silk 1986;
Dekel \& Woo 2003; Benson \et 2003),
and the mass-metallicity relation among galaxies (Larson 1974). Ablation of 
the gas in dwarf galaxies by winds from nearby galaxies may help flatten the 
faint-end of the luminosity function (Scannapieco, Thacker, \& Davis 2001).

Observations strongly support the idea that a larger fraction 
of the metals are removed from dwarf galaxies than massive galaxies.  
Metallicity measurements from X-ray spectra indicate the hot wind carries 
the heavy elements produced by the starburst population (Martin \et
 2002).  Temperature estimates for the hot gas, i.e. $kT_x$, 
vary little with galactic mass indicating a critical galaxy mass, 
$V_c \sim 130$\kms, below which most of the hot wind escapes (Martin 1999).
Retention of a larger fraction of the metals in more massive galaxies would 
produce an increasing effective yield with galaxy mass up to this scale
where the stellar yield is recovered (Larson 1974; Dekel \& Silk 1986; 
Vader 1987).  Although the X-ray data have some significant shortcomings --
including a small number of observed objects and a temperature biased
towards the densest regions of the hot plasma, the mass-metallicity relation
has been measured locally for a complete sample of galaxies.
The metallicities increase steeply
with mass over the range from $10^{8.5} - 10^{10.5}$\msun $h_{70}^{-2}$
but flatten at $10^{10.5}$\msun $h_{70}^{-2}$ (Tremonti \et 2004). The
mass scale of the flattening appears to be consistent with the inferred
mass scale of metal retention (Garnett 2002).  Despite the efficiency
of metal ejection, it is not clear whether the total mass loss rates which
are of the order of the SFR are sufficient to explain the measured 
faint-end slope of the galaxy luminosity function, which is
flatter than  that of the halo mass function (Somerville \& Primack 1999).

This paper finds an increase in the velocities of cold clouds with galactic 
mass suggesting feedback may play a more prominent role in massive galaxies
than previously thought, an idea which could be important for several
outstanding problems. First, wind velocities are quite important in 
determining how much of the IGM is polluted (Aguirre  2001). Fast winds from 
quasars may well deposit more energy into the IGM than starburst winds 
(Scannepieco \& Oh 2004), but they seem unlikely to dominate the metal transport.
Second, the galaxy luminosity function cuts off more steeply than the 
theoretical halo mass function at bright luminosities, and a feedback
mechanism physically distinct  from that in dwarf galaxies appears to
be required to explain both ends of the luminosity function. The mass scale 
separating these forms of feedback shows up not only in the mass-metallicity
relation and \x temperatures but also characterizes the galaxy mass
scale $\sim 3 \times 10^{10}$\msun\ where (1) surface-brightness stops rising 
with luminosity and mass and (2) less massive galaxies have stellar 
populations weighted towards younger ages (Blanton \et 2003; Kauffmann \et 
2003a,b). Any feedback processes invoked to explain these relations must do 
so without disrupting the Tully-Fisher and Faber-Jackson relations between
velocity, i.e. the depth of the gravitational potential, 
and stellar mass, $V \propto M_*^{1/4}$ (e.g. Bernardi \et 2003), which
show no obvious change at this scale (Zwaan \et 1995; Sprayberry \et 1995; 
Dale \et 1999).

Sample selection and the observational strategy are described in \S 2.
The kinematics and column densities of the nuclear ULIG spectra 
are presented in \S 3. Section~4 develops scaling relations between 
outflow properties and galaxy parameters, including the dynamical
state of the merger. The results are shown to challenge the standard
model of supernova-driven winds in \S 5, and the role of radiative
acceleration is discussed.
\S 6 summarizes the main results.
The second paper in this series (Paper II) describes the surprisingly
large spatial extent of \nad\ absorption, compares kinematics of
\nad\ and \Ha lines, and discusses the outflowing mass flux in
detail.

\section{Observations}

The spectra presented here extend \nad\ studies of dwarf 
starbursts (Schwartz \& Martin 2004) and luminous infrared galaxies 
(Heckman \et 2000) to ultraluminous galaxies (ULIGs). Targets were
chosen from the IRAS (60 micron) 2 Jy sample (Strauss \et 1992) 
to have bolometric luminosities greater than  $5 \times 10^{11}$\lsun,
a 60\um\ bump  $F_{\nu}^2 (60\um) > F_{\nu} (12\um) \times F_{\nu}(25\um)$,
and declination $\delta > -35\deg$.  Spectral observations of 41 of the
64 galaxies satisfying these criteria were obtained. The selection was
random unlike the Rupke \et (2002) sample which pre-selected galaxies with 
outflows from the 1~Jy survey.  The rather large redshift range, $z=0.042$ to 
$z=0.16$, reflects the paucity of ultraluminous galaxies locally. 

Echellete spectra were obtained 2000 September 19-20 and 2001 March 26-29
using ESI on Keck II. The nucleus, identified in R and K band images
(Murphy \et 1996), was placed on the slit.  The position angle was usually
chosen to include the second nucleus.  If the object could not be observed 
when the parallactic angle swept through the position angle defined by
the two nuclei, then an angle closer to parallactic was chosen.  The 
median seeing was 0\farcs8 FWHM (full width at half-maximum intensity),
which resolves the galactic disks.  The 1\asec slit provided a resolution 
of $\sim 70$\kms.  

The data were reduced using the echelle package in 
IRAF.\footnote{
	IRAF, the Image Reduction and Analysis Facility, is 
	written and supported by the IRAF programming group at 
	the National Optical Astronomy Observatories (NOAO).
 	NOAO is operated by the Association of Universities for 
	Research in Astronomy (AURA), Inc. under cooperative
        agreement with the National Science Foundation.}
Fixed pattern noise was removed from the CCD frames using bias frames
and internal quartz illumination frames. 
Fitted arc lamp exposures (Cu+Ar+Xe) provided a dispersion solution
accurate to $\sim 0.15\angs$.
The nuclear spectra were traced and extracted, and a sky spectrum 
extracted from the same slit was interactively scaled and subtracted.
The continuum was fitted and normalized to unity in each order.  
Velocities quoted in the paper are converted to the rest frame and 
given relative to the Local Standard of Rest.

The primary uncertainty in the outflow velocity of the cool gas
is the systemic velocity of the galaxy. 
This paper focuses on the {\em CO-sample} which contains 15 ULIGs with 
CO velocities from Solomon et al. (1997) and 3 ULIGs with CO
velocities from Dr. Aaron Evans (pvt. comm.). The stellar 
velocities derived from the Mg~I lines are fully consistent with the CO 
velocities although the uncertainties were typically larger. The CO emission 
was chosen mainly for its insensitivity to extinction and the following
dynamical consideration. In galaxy -- galaxy mergers, 
the interstellar gas sinks toward the dynamical center of the merged 
galaxies as the orbital angular momentum is shed at large radii. The 
central gas surface density grows to more than 10 times the maximum in the 
Milky Way, and the ISM  is almost entirely molecular (Sanders \et 1988a).
The measured CO velocity therefore provides a better description
of the systemic velocity than either HI 21-cm emission, which comes
from a larger region, or emission lines from HII regions. For the
CO-selected sample described in this paper, the typical accuracy of 
the CO velocities is $\sim 20$\kms (Sanders \et 1988b).
In the CO-sample, the minimum luminosity is $6.5 \times 10^{11}$\lsun, 
and the implied star formation rates are several hundred \msunyr\ as shown
in Table~\ref{tab:sample}. Four of the galaxies are members of the well-studied 
Bright Galaxy Sample (Sanders \et 1988b; Sanders \et 2003).

\section{Results}

Figure~\ref{fig:nad_pro} shows the \nad\ absorption lines
in the nuclear spectra. Three features are immediately obvious.
First, the doublet lines which are separated by
300\kms\ are blended together in all the spectra, so the line
widths are much larger than those observed in dwarf starbursts
and LIRGs. Second, the total \nad\ equivalent-widths, see 
Table~\ref{tab:PROFILE} are large -- ranging from 1 to 10\angs. 
And, third, the  lines are not black at the line center.  
In this section, we determine the contribution of interstellar 
absorption to these line profiles, describe
the kinematics of the absorbing clouds, and estimate the \nad\
column density.

%The velocity resolution, 70\kms, is more than sufficient to resolve the 
%doublet, but strong Doppler broadening blends the \line5889.95 
%and \line5895.92 lines together.
% however, so the absorbing material
% cannot completely cover the light source if the lines are
% optically thick.  

\subsection{Distinguishing Interstellar and Stellar Na~I}

Interstellar absorption lines in galactic spectra are formed against
a continuum which contains photospheric features. Since the
alkali metals are mostly neutral in the atmospheres of late-type stars, 
such stars present prominent \nad\ absorption lines. The strength of 
{\it excited} photospheric features, which have no interstellar counterpart, 
such as the Mg~I \line5167.32, \line5172.68, \line5183.60 triplet 
constrain the contribution of stellar \nad\ absorption however.
In ESI spectra of AFGK dwarf and giant stars, taken in the same
manner as the galaxy data, the strength of the NaD doublet relative
to the Mg~I triplet is well-described by the empirical relation 
$EW_{NaD} = 1/3 EW_{MgI}$.

In galaxy spectra, the Mg~I lines are blended with other photospheric lines,
particularly an Fe~I line, because they are broadened by 200-300\kms\ FWHM.
Figure~\ref{fig:namg} compares the total Na~I equivalent width to
the measured equivalent width of the entire Mg~I+Fe~I complex,
rest-frame bandpass from $\lambda  5162 \rm{ ~to~} \lambda 5189$,
in the nuclear ULIG spectra. Excited photospheric features are  strongest
in \i00153+5454, \i00188-0856, \i17208-0014, \i20087-0308, and \i23365+3604;
but these galaxies also present \nad\ absorption much stronger than that 
predicted by the stellar locus. Although the \nad\ strength for 
\i00262+4251 exceeds the predicted stellar equivalent width by only one 
standard deviation of the measurement errors, the significant Doppler
shift of the lines suggest an interstellar component is present.
In \fig\ref{fig:namg}, the upper limit for 
\i19458+0944 is not very restrictive, but the upper limit for the other null
detection, \i16487+5447, allows only a stellar component.

Table~\ref{tab:PROFILE} lists the total NaD equivalent width (in the 
rest-frame $\lambda  5874 \rm{ ~to~} \lambda 5899$ bandpass) and
the fraction inferred to be stellar from the  scaling
between stellar \nad\ and Mg~I. The  stellar contribution to the NaD equivalent 
width is typically $\sim 10\%$ and will be ignored in the remainder of
this paper.

\subsection{Interstellar Gas Kinematics} 

Since the stellar \nad\ absorption is negligible,
the line profiles describe the kinematics of interstellar gas clouds 
and, if a resolution element is optically thin, the \nad\ column density 
at a particular velocity. Doppler shifted absorption components
are a robust signature of outflow (for a blueshift) since the absorbing
gas must lie on the near side of the galaxy.  One obtains no direct 
information about the location of the absorbing gas along the sightline,
however. As discussed in \S2, the systemic velocities for the CO-subsample
are well determined. The systematic analysis of the \nad\ absorption presented 
here draws the robust conclusion that nearly all ULIGs drive fast outflows. 
For any particular ULIG, the interpretation may be complicated by absorption 
from tidal debris (e.g. Norman \et 1996; Hibbard, Vacca, \& Yun 2000).
Tidal debris, in contrast to starburst-driven outflows, subtend a 
small solid angle (L. Hernquist, pvt comm), so they are unlikely  to be
the dominant source of the blue-shifted absorption lines.

\subsubsection{Doppler Velocities}

The spectral region around \nad\ $\lambda \lambda 5890, 96$ is
plotted in \fig~\ref{fig:nad_pro}. Comparison of the systemic velocity,
marked by a doublet, to the line profiles reveals that much of the absorbing 
gas is Doppler shifted to wavelengths shorter than the that of the rest-frame
doublet. To describe this {\it dynamic component}, a second doublet was fitted 
while the velocity of the systemic component was held fixed.  Although the
best fits, which are optically thick, are summarized in Table~\ref{tab:tableFIT},
another set of models with the ratio of the line strengths held
equal to the ratio of the oscillator strengths (two-to-one) is provided in
Table~\ref{tab:Nthin}. Comparison of the velocities in these two tables
illustrates the maximum systematic uncertainties in the measurements.

Although the fully resolved line profiles might be more complex than
these simple Gaussian fits assume, the fitted velocities of the dynamic 
absorption component, $v_B$ in Tables~\ref{tab:tableFIT} and \ref{tab:Nthin}, 
provide a fair estimate of the mass-weighted outflow velocity.  
The average fitted outflow velocity is  $330 \pm 100$\kms. The 
uncertainty in $v_B$ is only a few \kms\ when the profiles show multiple local 
minima. For example, superposition of the shorter wavelength line from the 
systemic component and the longer wavelength member of the dynamic component 
creates three local minima in \i08030+5243 and \i17208-0014. Two minima and 
a strong blueshifted line wing require two blueshifted components in 
\i10565+2448 and \i18368+3549,  and three components were fitted with one
component held at the systemic velocity. Fitting smooth, blended line profiles, 
on the other hand, does not yield a unique solution, and these  systems
have uncertainties up to $\pm 70$\kms in $v_B$ in Table~\ref{tab:tableFIT}. 
Selection of a low doublet ratio for the systemic component systematically
lowers the fitted velocity of the dynamic component.\footnote{
  When  $D_{sys}$ is varied from unity to its optically thin limit,
  $D_{sys} = 1.98$, the systemic component absorbs more flux in the center 
  of the total line profile; and, as a consequence, the velocity of the dynamic 
  component shifts to larger outflow velocities.}
Comparison of $v_B$ in the optically thin limit,  Table~\ref{tab:Nthin},
and the optically thick limit, Table~\ref{tab:tableFIT}, shows the magnitude 
of the shift is generally negligible. Two exceptions are \i00262+4251, where 
the shift is  100\kms, and \i10494+4424, where the discrepancy is $110 \kms$.
In both cases, the model in Table~\ref{tab:Nthin}, which has  the saturated
systemic component, is favored.\footnote{
                  In the \i00262+4251 fit, the fit statistic is not 
		  acceptable for the unsaturated 
		  model, so the -293\kms of the saturated model best 
		  describes the outflow. In \i10494+4424,
                  the saturated model  describes the two
		  minima better 
		  than the unsaturated model.}
Systematic uncertainties in $v_B$ caused by the choice of the doublet
ratio are therefore insignificant compared to the magnitude of the outflow
velocities.

The maximum outflow velocities were measured where the \nad\ line profile 
intersects the continuum. These velocities presumably reflect the terminal 
cloud velocity significant distances from the starburst.  These
measurements are not influenced by the details of the line fitting, so
they are tabulated with the basic measurements in Table~\ref{tab:PROFILE}.
The mean terminal velocity is -750\kms, which is a few times larger than the
galactic rotation speeds. This coincidence is particularly interesting since
the escape velocity from the galactic halo is also a few times the rotation 
speed. The physical meaning of this correlation is explored further in
\S 5.

The He~I 5876 emission -- present in all spectra except
\i00188-0856, \i03158+4227, \i18368+3549, and \i20087-0308 --
could cause the true terminal velocity to be systematically underestimated.
At velocities greater than  729\kms, absorption from the 5889 Na line 
is observed at the same wavelength as systemic He~I 5876 emission.
Three factors indicate this coincidence does not bias the
measured terminal velocity. First, the apparent strength of the
He~I 5876 emission line relative to the upper limits on the
He~I 4471 line (another triplet with the same lower state as He~I 5876) and 
He~I 6678 (the n=3 to n=2 singlet transition) indicates
little, if any, of the He~I 5876 emission is
cancelled by very high velocity Na~D absorption.  Second,
the \nad\ absorption extends further along the slit than the He~I emission
in many cases, and Paper II shows similar terminal velocities are
measured all along the slit. Finally, the K~I $\lambda \lambda 7665, 99$
doublet, although a weaker line than \nad, would be expected to trace
the same neutral gas.  The doublet spacing is 1327\kms\ so these lines are 
not blended. Figure~\ref{fig:KI_NaI} shows that the K~I absorption 
profile closely follow the shape of the \nad\ profile in \i10565+2448 and 
\i20087-0308. Although the K~I absorption profile for \i15245+1019 reveals an
absorption line blueward of the \nad\ terminal velocity, this
candidate high velocity component is not detected in the K~I $\lambda 7699$
line (not pictured) and therefore is not the K~I $\lambda 7696$ line.
The K~I lines were not detected in the nuclear spectrum for \i17208-0014,
and residuals from subtraction of strong OH sky lines prevent an accurate 
measurement of the K~I terminal velocity in \i18368+3549.
Hence, when  the K~I doublet was detected, it presented the same terminal 
velocity as \nad.

In summary, among the 18 galaxies in the CO-subsample, 15 present neutral 
gas outflows.
The galaxies \i16487+5447 and \i19458+0944 show no (or weak)  interstellar 
\nad\ absorption. Only the \i08030+5243 line profile shows a redshifted 
component.\footnote{
  The absorbing gas in \i08030+5243 appears to be falling toward the 
  CO-defined systemic velocity at 250\kms. 
  Since no redshifted flows are detected in the other 17 galaxies,
  and an outflow from one galaxy can be detected in absorption against the 
  continuum from another galaxy (see Paper II), one wonders 
  if the orientation of the sightline toward \i08030+5243 is unusual 
  or special. The object 10\asec southeast of \i08030+5243, however, has been 
  classified as a foreground star based on its surface brightness profile, so
  there is no obvious companion galaxy  (Murphy \et 1996).}
This outflow fraction of at least 80\% in ULIGs, which is higher than the 
outflow fraction in luminous infrared galaxies and (Heckman \et 2000) 
dwarf starbursts (Schwartz \& Martin 2004). The most luminous LIGs also
presented a higher outflow fraction than lower luminosity LIGs, but
the significance of the trend was unclear because the two subsamples
were selected differently. The new ULIG data establish a trend of increasing
outflow fraction with luminosity. Indeed, all ULIGs may have outflows, and 
this hypothesis is explored further in \S 4.3.

% The KI lines for \i15245+1019 fall 
% in a band gap where the sky can be cleanly subtracted. 

\subsubsection{Line Widths} 

In Table~\ref{tab:tableFIT}, the average line width of the dynamic component 
is  $<\Delta v_B> = 320\pm120$\kms FWHM, much larger than the thermal 
velocity dispersion of warm neutral gas in galaxies. For comparison,
an average sightline through the Milky Way disk detects 6 clouds per kpc 
in \nad\  -- each with velocity dispersion $\sim 8$\kms (Spitzer 1968).
Those LIGs with interstellar-dominated \nad\ absorption also present very
broad lines (Heckman \et 2000).  While turbulent motion, particularly
at the interfaces between the hot wind and the disk (e.g. Figure~11 in
Heckman \et 2000), is expected to broaden the lines, numerical simulations
of bipolar outflows appear to have difficulty producing the large observed 
velocity range (D. Strickland, pvt comm). 

To produce a width of 300-400\kms, a sightline may need to intersect multiple 
turbulent sheets or shells, each with distinct bulk motion. Higher resolution 
spectra (6\kms in  \fig\ 1 of Schwartz \& Martin 2004) support this claim. 
Along a sightline into the center of M82, the broad \nad\ doublet, 236\kms\ 
FWHM, is resolved into 5 distinct velocity components, each of width 
25 - 79\kms.  And, although the \nad\ profile remains smooth in \n1614,
the more optically thin K~I lines do break up into multiple components. 
An important test of dynamical models is to determine, presumably using 
numerical simulations, whether a more realistic spatial distribution of star 
formation can generate multiple outflow shells and a broad enough velocity 
distribution.

% relaxed population in SSC LIGs -- low FWHM -- not sampling full potential

The fitted systemic \nad\ component probes a mixture of stellar
absorption and interstellar absorption at the systemic velocity.
The formal uncertainties for the widths of the systemic component are 
significantly larger than those for the dynamic component because the systemic 
component is generally weak.  The average width of the systemic component
$\Delta v_{sys}$ is $400 \pm 140$\kms FWHM corresponds to a line-of-sight 
velocity dispersion, $\sigma = 170 \pm 60$\kms.
These widths are substantially larger than those measured for
stellar-dominated LIGs (Heckman \et 2000) but lie at the low end
of the range measured in ULIG spectra from stellar lines   $\sigma_* =
140 - 290$\kms\ (Genzel \et 2001).  The \nad\ sample presented in this paper
has three galaxies in common with the Genzel \et sample.
In \i17208-0014, the stellar velocity dispersion, $\sigma_* = 229\pm15$ \kms\
and $v_* = 110$\kms, is subtantially larger than that of the systemic \nad\ 
component, $\sigma_{sys} = FWHM / 2.35 = 108$\kms. However, in \i23365+3604 
the systemic \nad component,  $\sigma_{sys}(\nad) \approx 259$\kms\ is
much larger than the stellar velocity dispersion, $\sigma_*  = 145\pm15$ \kms\
and $v_* = 15$\kms.
The  widths are similar in  \i20087-0308 -- $\sigma_* = 219\pm14$ \kms,
$v_* = 50$\kms, and $v_{sys}(\nad) =  203\kms$; but this may be largely
coincidental considering that the uncertainty in $\Delta v_{sys}$ exceeds 
100\kms. It is not surprising that the fitted \nad\ systemic components do 
not appear to be accurate measures of the stellar velocity dispersion --
their strength suggests a blend of stellar and interstellar absorption at
the systemic velocity.

% The mean velocity width of the dynamic component, $\Delta V_b = 340 \pm 
% 130$\kms, is enormous compared to the thermal line width for cold, 100~K 
% clouds,  ${\rm FWHM~} \nad \approx 0.5 ~T_2^{0.5}$\kms. 

% MW radii $\sim 7 $~pc (Spitzer 1968)

% USE THIS IF YOU TRY TO INTERPRET THE VELOCITY WIDTHS
%
% The width of the systemic component
% $is well constrained by spectra which show a lot of structure
% such as \i17208-0014. The fitted FWHM is $254 \pm 6$\kms\ 
% in the optically thick model and changes little in the extreme
% limit of the optically thin model.  In contrast, $\Delta v_{sys}$
% is essentially unconstrained in \i19297-0406 and \i00262+4251 which
% have smooth line profiles. 

% DON'T NEED ANOTHER DESCRIPTION OF THE MAX FLOW VELOCITIES
%
% The fitted width of the profile also provides another characterization
% of the maximum flow velocities. The sum of the half width  and the
% velocity offset of the line center indicate
% the clouds reach an average velocity of {\bf FIX  $457 \pm 173 $\kms.}
% This quantity can be used in place of the terminal velocities in
% column 3 of Table~\ref{tab:PROFILE} for some applications.

% The wing of the fitted blue component gives higher 
% terminal velocities in a few cases. The discrepancies
% indicate the magnitude of our uncertainty about the terminal
% velocity in any individual system.  The maximum discrepancy
% is 300\kms.  This detail affects the deconvolution of
% the line profiles.  

\subsection{\nad\ Columns}

Measurements of the \nad\ column density in the dynamic component 
provide a basis for estimating the mass flux of outflowing material 
per unit area. These measurements differ from absorption measurements using
quasars in that the continuum is extended (slit width subtends 
$\sim 1 - 2$~kpc of galaxy) and extremely dusty sightlines are
excluded from the average.\footnote{For gas and dust  mixed in roughly the 
            Galactic ratio, sightlines with $N(\nad) > 1.7 
	    \times 10^{22}\col$ would have visual extinction 
	    $A_V > 10$ completely obscuring the continuum source.} 
In principle, measurements of the K~I lines are more sensitive
to column density than the \nad\ lines, which are shown to suffer
from saturation; but subtraction of bright sky lines near K~I
generally leaves large uncertainties about the K~I line strength.

% The oscillator strengths are lower than for \nad\, and
% the solar abundance ratio of Na/K is $\sim 15$, the KI lines
% are likely optically thin. 

\subsubsection{Fitted Models}

The inferred \nad\ column is only directly proportional to the measured \nad\ 
equivalent width if the absorbing material is optically thin. 
Assuming optically thin clouds, the derived \nad\ columns range from 
$\sim 10^{12}$\col\ to a few times $10^{13}$\col, see Table~\ref{tab:Nthin}. 
Since the optical depth of the stronger doublet member, $\nad\ \lambda 5890$, 
reaches unity at a column of $N(\nad) \approx 3.6 \times 10^{13}\col\ 
(v / 100 {\rm ~km s~}^{-1})^2$, the \nad\ lines are not guaranteed to be 
optically thin. The columns in Table~\ref{tab:Nthin} are good lower limits.
The fitted doublet ratios, $D \equiv EW(\lambda 5890) / 
EW(\lambda 5896)$, in Table~\ref{tab:tableFIT} illustrate how much 
larger the true \nad\ column might be.\footnote{
          The models adopt an 
          optically thick {\it systemic} absorption component. 
          The strength of the $\lambda 5896$ line in the systemic component 
	  is limited by the overall line profile, so the effect of increasing 
	  the doublet ratio from 1.0 to 1.98 is to strengthen the \line 5890 
	  line. Since the Doppler shift of the dynamic component superimposes 
	  the $\lambda 5896$ line of the dynamic component on the systemic 
	  $\lambda 5890$ line, the fitted \line 5896 line of the dynamic 
	  component becomes weaker and the doublet ratio of the dynamic 
	  component is larger (i.e. lower column).} 
The fit statistic was improved for every galaxy when the doublet ratio
was allowed to decrease from the optically thin limit, $D =  1.98$.
Figure~\ref{fig:duo} shows that the thick models do a better job of fitting 
the structure in the line profiles. Optically
thin fits cannot be ruled out for the smooth, featureless profiles in
\i00262+4251 and \i20087-0308,  but there is no obvious reason to expect these 
systems to be optically thin when the better constrained fits all indicate
optically thick outflows toward the nucleus.

% The absorption is split roughly evenly between the systemic and dynamic
% components in \i10494+4424, but 

\subsubsection{Cloud Covering Factor}

\begin{figure}[h]
      {\includegraphics[scale=0.7,angle=0,clip=true]{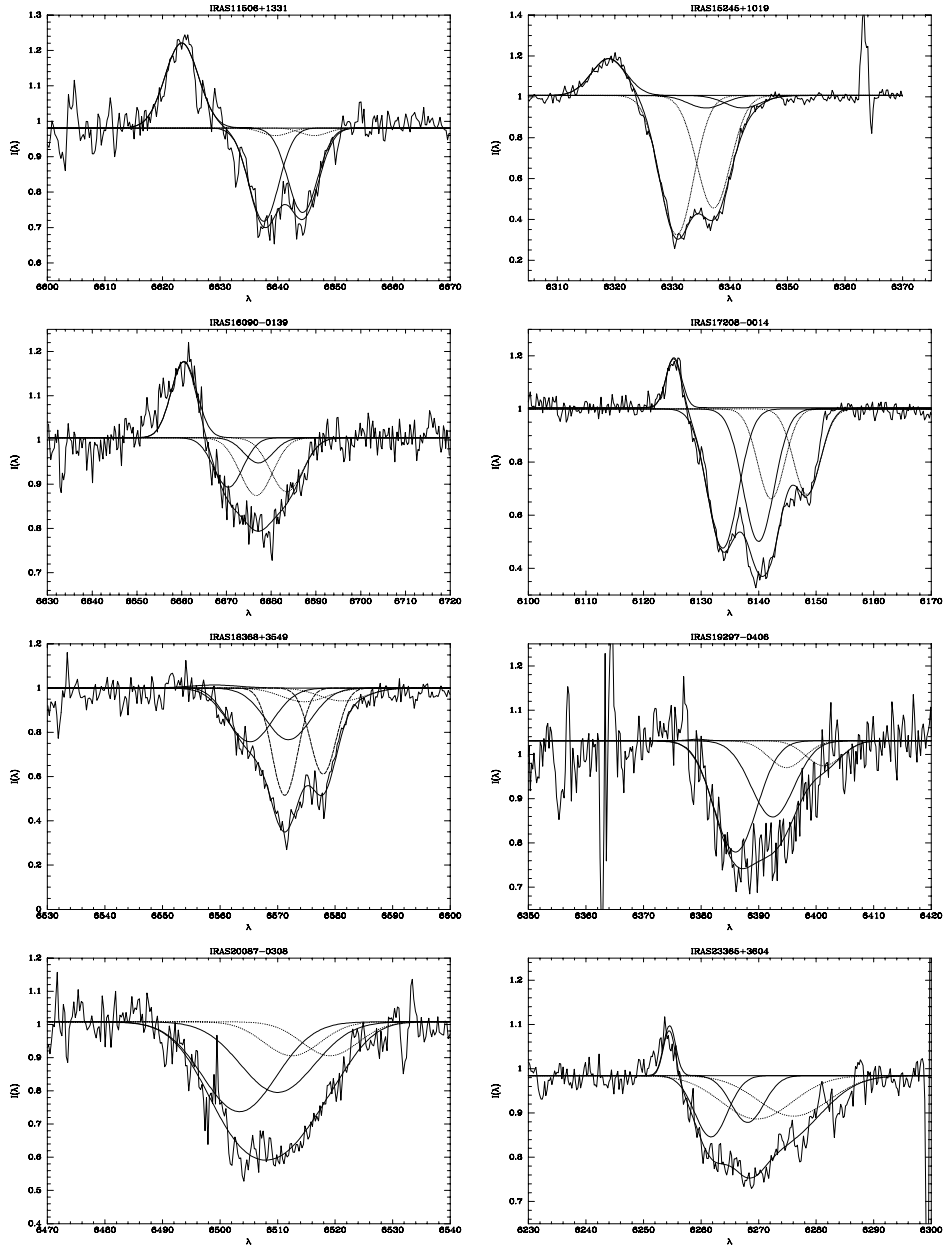}
	\hfill}
        \caption{(b) The ESI spectra around  the \nad\ $\lambda 5890, 96$ 
	  doublet. The emission line is He~I $\lambda 5876$. 
	  No \nad\ absorption was detected in \i16487+5447 and \i19458+0944.
	  The dotted line shows a doublet at the systemic velocity,
	  which was determined from observations of the molecular
	  gas. The doublet ratio of the  systemic component is fixed
	  at $EW(5890)/EW(5896) = 1.06$ here.  The Doppler shift
	  and doublet ratio of the {\it dynamic component}, shown
	  by the thin solid line, were fitted. The heavy solid line
	  shows the sum of the dynamic component, systemic component, and 
	  the He~I emission line.  All but one galaxy presents a highly
	  blueshifted dynamic component.
}
\label{fig:nad_pro} \end{figure}

\begin{figure}[h]
      {\includegraphics[scale=1.0,angle=0,clip=true]{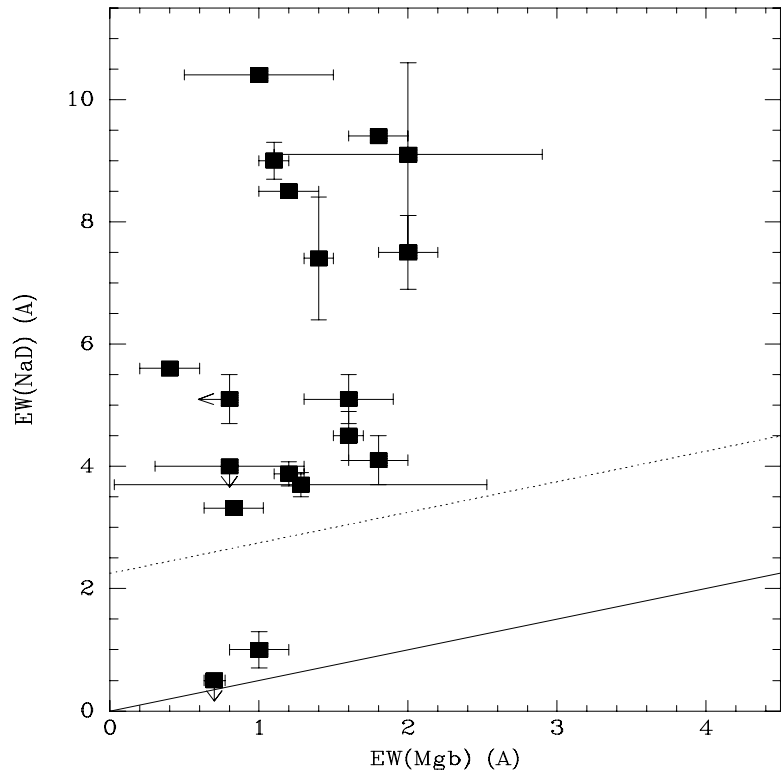}
	\hfill}
        \caption{Measured Na~D absorption equivalent width versus 
	  measured Mg~I absorption equivalent width in ULIGs. The
	  solid line is a fit to stellar spectra (see \S 3.1 of text),
	  and the dotted line illustrates typical values in 
	  elliptical galaxies (Bica \et 1991).}
\label{fig:namg} \end{figure}

\begin{figure}[h]
      {\includegraphics[scale=1.0,angle=-90,clip=true]{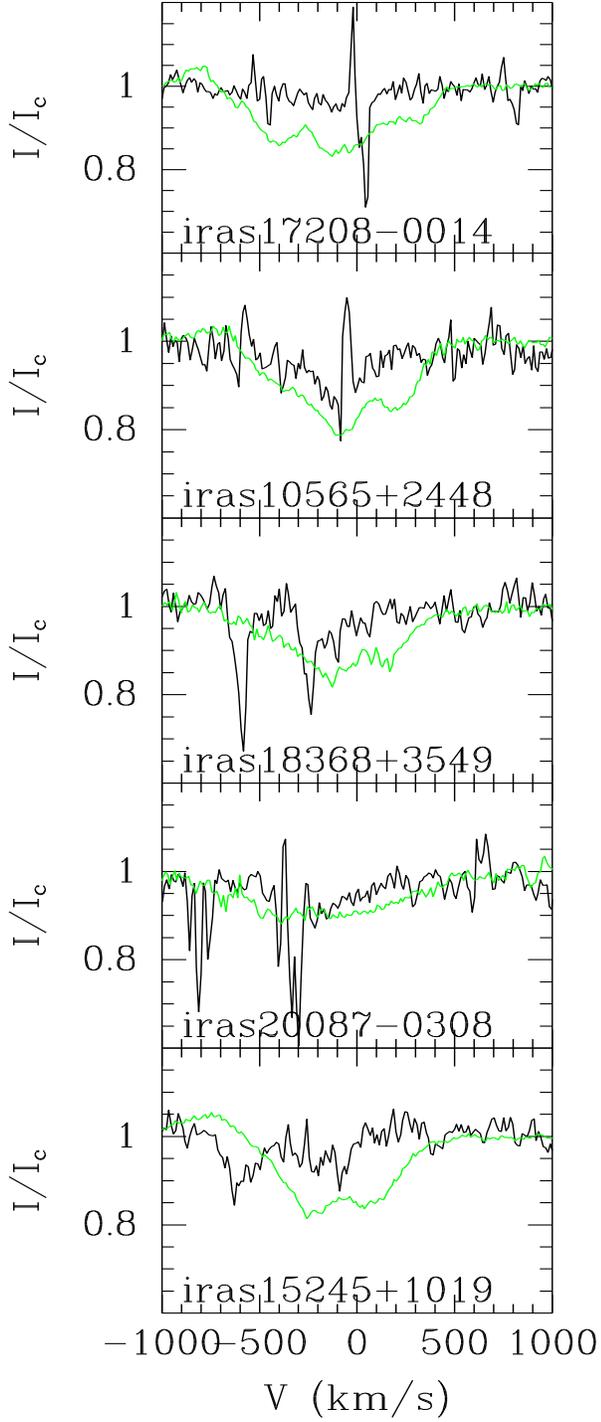}
	\hfill}
        \caption{The K~I 7664.9 profile for the five galaxies with the
	  strongest interstellar absorption. The weaker member of
	  the doublet is at $+1327$\kms\ (not shown). The \nad\ doublet,
	  arbitrarily scaled, is overlaid.    The profiles deviate, as they 
	  must, redward of the systemic velocity where the \nad\ 5890 appears. 
	  The spikes in the K~I spectra are residuals from the subtraction
	  of bright night sky lines and should be ignored. The K~I 7664.9 
	  absorption profile is 
	  generally similar in shape to the \nad\ 5889 	profile but has 
	  poor signal-to-noise ratio.}
\label{fig:KI_NaI} \end{figure}

\begin{figure}
      {\includegraphics[scale=0.35,angle=-90,clip=true]{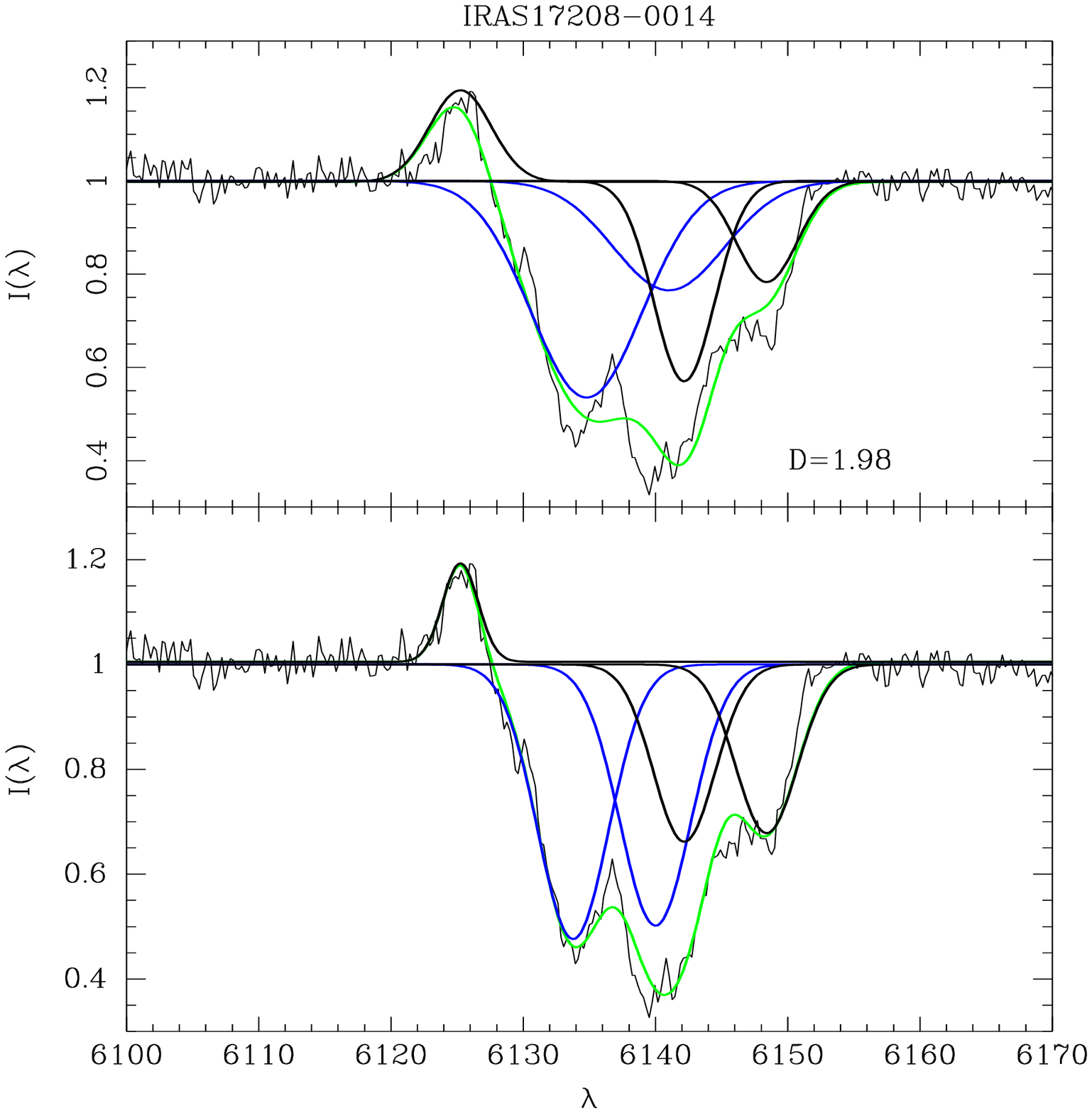} 
        \includegraphics[scale=0.35,angle=-90,clip=true]{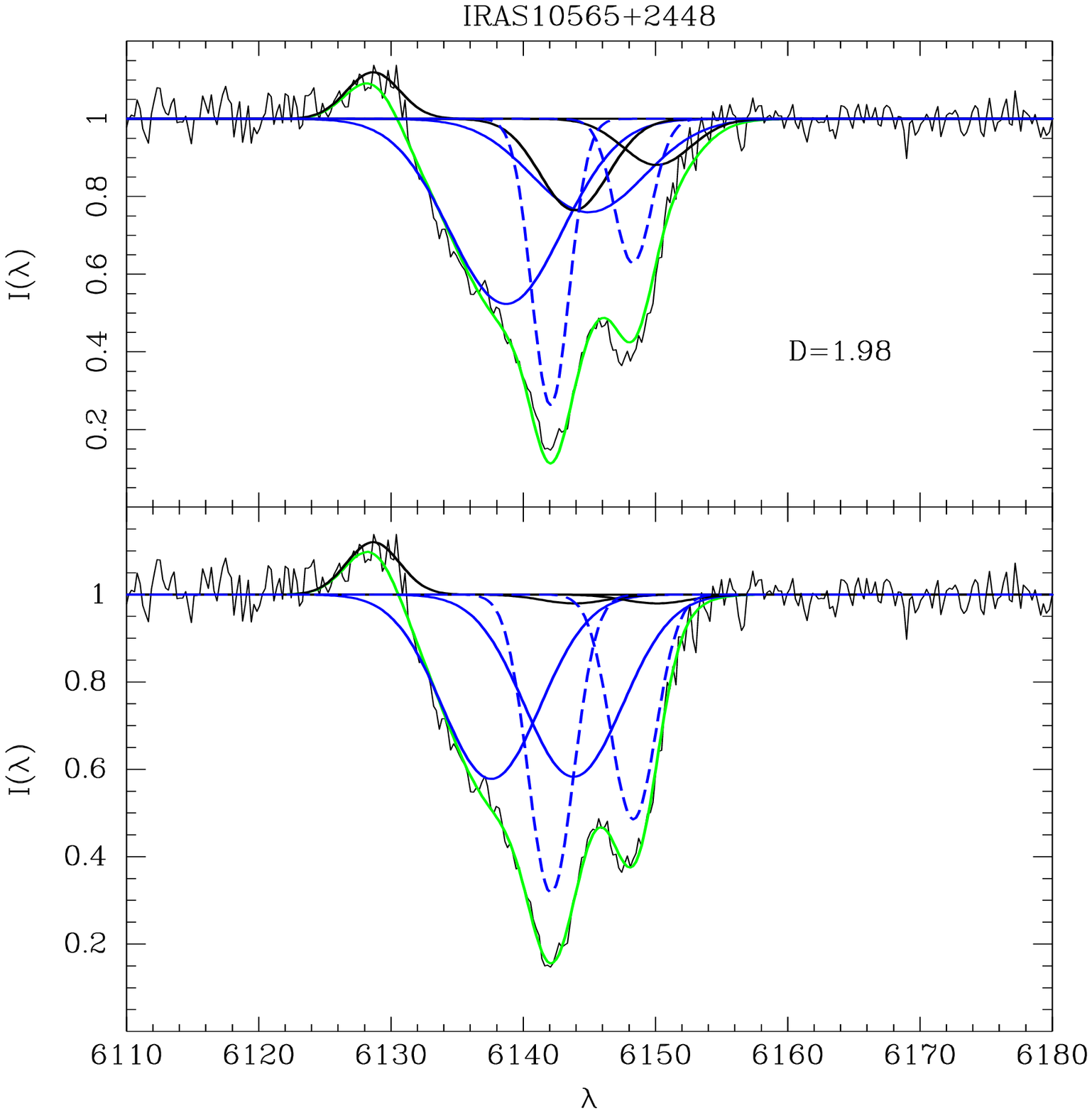}
	 \includegraphics[scale=0.35,angle=-90,clip=true]{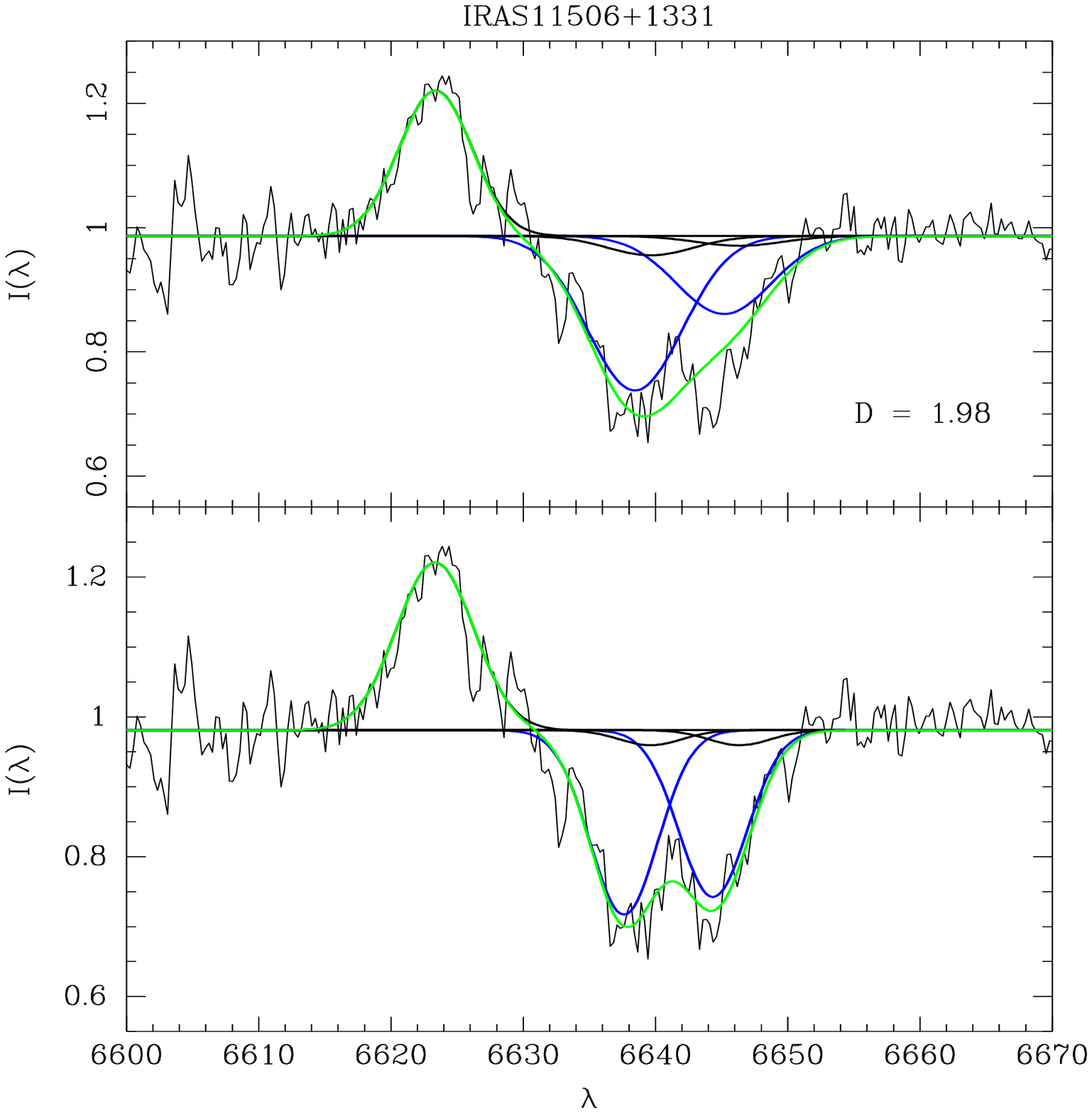} 
          \hfill}
        \caption{Examples of fitted \nad\ absorption lines in the optically
	  thin (top panel) and thick (bottom panel) limits. The blue line
	  shows the fitted dynamic component, and the doublet at the 
	  systemic velocity is drawn in black. The green line shows the
	  sum of the model components. The optically 
	  thick models describe the local minima better than the optically
	  thin models.} 
\label{fig:duo} \end{figure}

Optically thick clouds that cover the continuum source would 
leave absorption lines completely black at the line centers.
None of the lines in \fig~\ref{fig:nad_pro} are black at line center.
The conclusion that the \nad\ lines are saturated in many, if not
all, of these galaxies directly implies that the absorbing clouds
do not completely cover the continuum source.  

In the limit of large optical depth, the cloud covering factor  $C_f$ is 
simply $ \approx 1 - I_{5890}$,  where $I_{5890}$ is the residual 
intensity in the \line 5890 line. The covering factors listed in
Table~\ref{tab:PROFILE}, which use the $I_{5890}$ value fitted in 
\fig~\ref{fig:nad_pro}, indicate that 25\% of the visible continuum 
source is covered by the clouds on average. The covering factors show
a relatively large range, however, from 10\% in
\i16090-0139  and \i00262+4251 up to nearly 70\% in \i15245+1019.

\begin{figure}
      {\includegraphics[scale=0.6,angle=0,clip=true] {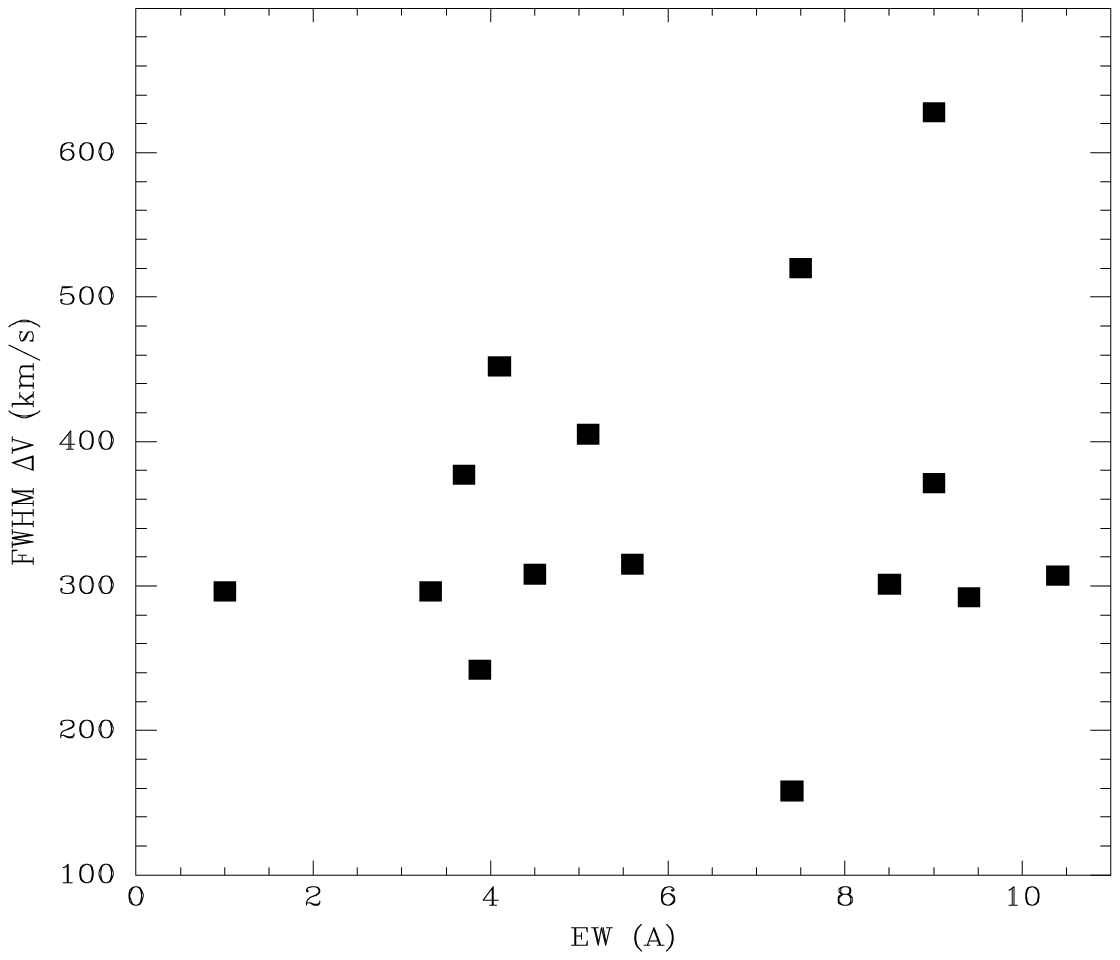} 
        \includegraphics[scale=0.6,angle=0,clip=true]{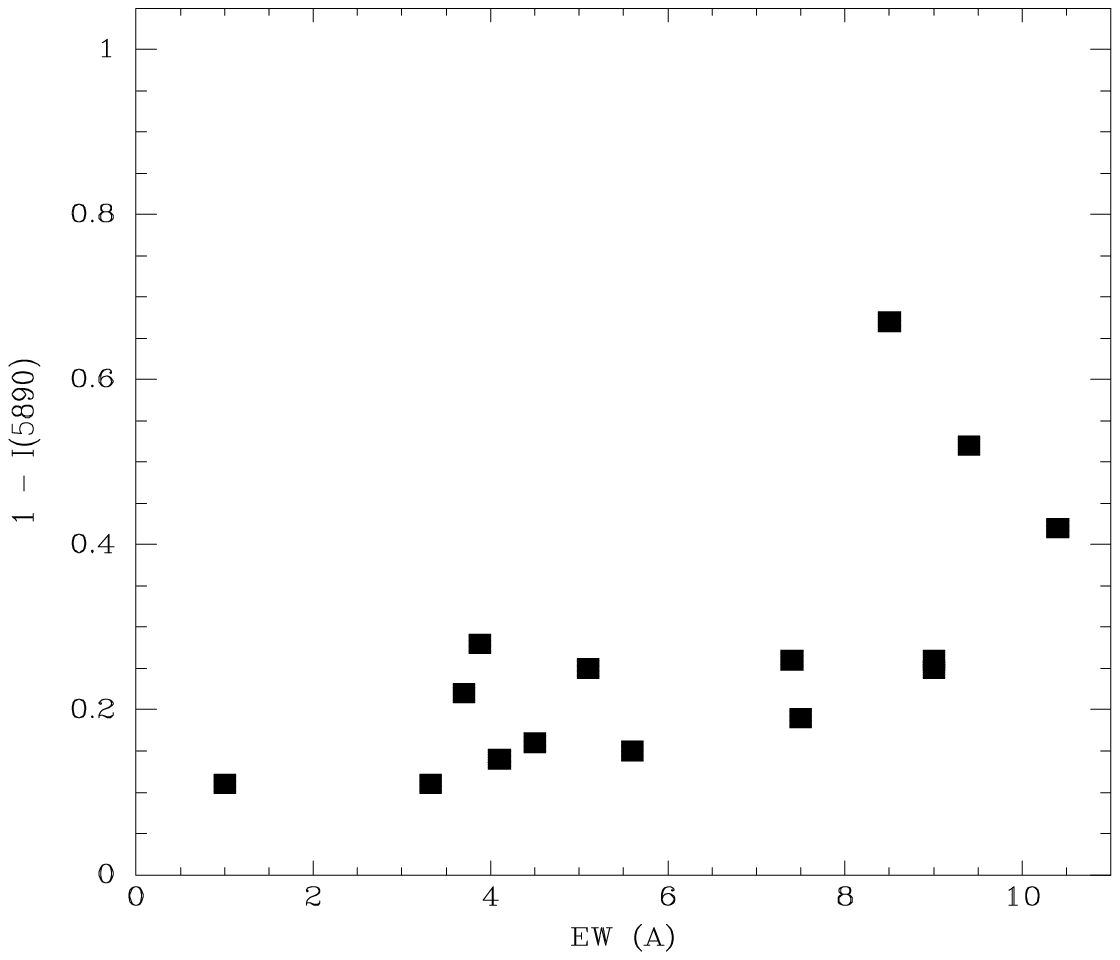}
	\hfill}
         \caption{(Left ) The \nad\ equivalent width shows no correlation 
	  with the velocity width of the line. (Right) Cloud covering 
	  factor versus \nad\ equivalent width in the dynamic component. 
	  The sightlines with the highest covering factors present the 
	  strongest absorption lines.}
\label{fig:dv_EW} \end{figure}

When optically thick gas only partially covers the continuum
source, the measured equivalent width will be insensitive to
the \nad\ column density and will be primarily determined
by the product of the covering factor and the line-of-sight
velocity dispersion.  Figure~\ref{fig:dv_EW} shows that the FWHM
of the dynamic component is not correlated with the total \nad\ 
equivalent width. It does not correlate with the EW of the dynamic
component either. The bottom panel of \fig~\ref{fig:dv_EW} shows that the
lowest equivalent width systems all present low covering factors,
and that large covering factors are only found in galaxies presenting  
very high \nad\ equivalent width.  A large covering factor of cool, 
outflowing gas may partially reflect the viewing angle, but the analysis
of \S 4.4 indicates temporal variations are also important.

% (The total widths of the line profiles at zero
% intensity are larger on average in the highest EW systems.)

\subsubsection{Conclusions about Column Densities}

Seven of the ULIGs
show extremely strong NaD equivalent width ($EW_o > 7\angs$); they are 
\i00188-0856, \i08030+5243, \i10565+2448, \i15245+1019, 
\i17208-0014, \i18368+3549, and \i20087-0308. The
optically thin models generally give larger \nad\ columns in 
the galaxies with larger \nad\ equivalent widths. The association
is not exactly one-to-one however. For example is \i15245+1019
is 5th in total EW but 3rd in total column because the particularly large 
Doppler shifts leave very little absorbed flux near the systemic
velocity. Hence, the fraction of the total equivalent width
provided by the systemic component is particularly low in
this galaxy.  The relative
 equivalent widths of the dynamic and systemic components
 can be ambiguous  when the profile is smooth.  Among the 7 largest 
 equivalent width systems, this happens in \i20087-0308 and
\i00188-0856 where fits with more absorption at $v_{sys}$
are clearly allowed.

Some of the profiles absolutely require very large fitted columns.  
Focusing on the systems with the largest equivalent width again,
one finds that the systemic component has to be very weak
in \i10565-2448, \i18368+3549, and \i15245+1019. Similarly,
the systemic component is strong but maxed out in the
fits to \i17208-0014 and \i08030+5243.  In summary, freeing
the doublet ratio raises the \nad\ column  to $N(\nad)
\sim 10^{14}$\col\ to $10^{15}$\col\ in galaxies for which the
optically thin model indicated large columns. The correction, however,
has little impact on the columns in weaker systems,$N(\nad)
\times 10^{12}$\col\ since it is not linear and therefore increases the 
range of measured \nad\ column densities.

% Two caveats should be considered when comparing
% the relative columns of the dynamic and systemic components.
% As previously explained, 
% systemic component is optically thick, $D_{sys} \approx 1$, rather
% than thin, $D_{sys} \approx 1.98$, the fitted column density
% of the dynamic component increases. One can easily see why by
% recognizing that the wavelength of the red
% member of the dynamic component is similar to the wavelength of
% the blue member of the systemic component, so decreasing the latter
% allows the former to be stronger.  The strength of the blue
% member of the dynamic component and the red member of the 
% systemic component are, in contrast, fixed by the overall
% shape of the observed line profile. 

\section{Discussion}

The absorption measurements presented here trace gas clouds with a significant 
neutral component.  As discussed above, the absorbing material is thought to be
disk gas entrained in a hot supernovae heated outflow. The X-ray 
observations are biased by conditions in the highest density regions of hot 
component and do not necessarily reveal the hottest (lowest density) regions
of the hot outflow (Strickland \& Stevens 2000), but they do show that
most of the shock heated mass can escape from dwarf galaxies with 
velocities less than 130 \kms  (Martin 1999; Martin 2004). More
of the shock heated gas remains bound in larger galaxies, and this
differential feedback contributes to the flattening of the slope of the
faint-end of the luminosity function.  No direct measurements of the 
kinematics are  currently possible in the hot phase. Since the clouds
absorbing in \nad\ are presumably embedded in this medium, it is possible
to learn about the wind fluid itself via the motions of the clouds.

\subsection{Outflow Mass}

The inferred H column associated with a sightline depends on
the chemical abundance, dust depletion, and ionization state
of the gas.  Assuming solar abundances in the ULIGs, the
H column is
\begin{eqnarray}
N_H = 4.90 \times 10^{20} ~{\rm cm}^{-2} 
\left( \frac{N_{NaI}}{10^{14} {\rm ~cm}^{-2}} \right)
\left( \frac{d_{Na}}{10} \right) \times \nonumber \\
\left( \frac {N(Na)}{N(NaI)} \right).
\end{eqnarray}
Heckman \et (2000) argue, based on parameters of Galactic
clouds, that the depletion of Na onto dust grains,
$d_{Na} = (N(Na)_{gas} + N(Na)_{dust}) / N(Na)_{gas}$, is
likely a factor $\sim 10$, and  the ionization
correction, e.g. $ N(Na)/N(NaI)$, could be as much as a 
factor of 3 to 10.

The nuclear sightlines sample a small region of each galactic halo,
so extrapolating these column densities to masses is very sensitive to
the chosen area.

Observations of the absorption-line profiles at multiple positions
across the galactic disks of ULIGs,  presented in Paper II, indicate
that some outflows cover more than 10~kpc of the host galaxy.
Taking a fiducial radius of $\sim 5$~kpc for purposes of 
illustration here,  the mass of clouds, $M \sim N_H A \mu m_p$ or 
\begin{eqnarray}
M_c \sim 1.08 \times 10^8 \msun \left( \frac{N_{Na}}{10^{14} {\rm ~cm}^{-2}} \right)
\left( \frac{d_{Na}}{10} \right) \times \nonumber \\
\left (\frac{N(Na)}{N(NaI)}\right)  
 \left( \frac{R}{5 ~{\rm kpc}}\right)^2.
\end{eqnarray}

are applied.  
The kinetic energy carried by the
cold clouds, $ 1/2 M v^2$, is
\begin{eqnarray}
E \sim 1.71 \times 10^{56}~{\rm ergs} 
\left( \frac{N_{NaI}}{10^{14}\col} \right)
\left( \frac{d_{Na}}{10} \right) \times \nonumber \\
\left( \frac{N(Na)}{N(NaI)}\right) 
 \left( \frac{R}{5 ~{\rm kpc}}\right)^2    \nonumber \\
 \left( \frac{v}{400 {\rm ~km~s}^{-1}}\right)^2.
\end{eqnarray}
The enormity of these quantities is perhaps most appreciated when
the mass loss rate, $\dot{M_c} = \Omega r^2 \rho v= \Omega r \mu m_H N_H v$ or 
\begin{eqnarray}
\dot{M_c} = 
141 \msunyr \left(  \frac{N_H}{4.9 \times 10^{20}{\rm ~cm}^{-2} } \right) 
\left( \frac{R}{5 ~{\rm kpc}} \right) \times \nonumber \\
\left( \frac{v}{400 ~{\rm km~s}^{-1}} \right) 
 \left( \frac{\Omega}{4\pi} \right),
\end{eqnarray}
is compared to the SFR.
For example,  applying these expressions to \i10565+2448 indicates
a mass flux in cool gas of $\sim 300\msunyr$, which is similar to the SFR of
190\msunyr.  Paper II presents spatially resolved line profiles and
more detailed estimates of the mass loss rates, but this example illustrates
the potential importance of the cold outflows.  If they can be
sustained for even 10~Myr, several billion solar masses of interstellar
gas is transported out of the disk thereby starving the starburst for 
fuel.

\subsection{Variation of Outflow Velocity with Galactic Mass}

One of the principal motivations for this study and that of 
Schwartz \& Martin (2003) was to investigate how outflow velocities 
scale with galactic mass. Surprisingly, Heckman \et (2000) found no 
correlation between outflow velocity and galactic rotation speed,  
or starburst luminosity. Simple extrapolation of this result would imply
dwarf galaxies completely dominate intergalactic enrichment.  
High-resolution \nad\ spectra of seven 
dwarf starburst galaxies, however, revealed outflows in three sytems; but 
the velocities,  $-25$\kms to -35\kms, were much lower than those
in the LIGs (Schwartz \& Martin 2004).
Provided their luminosities are dominated by the starburst rather than a quasar,
the ULIGs have star formation rates from 190 - 750 \msunyr, which
(by definition) is higher than that for LIGs which drop down to about 10\msunyr.
Outflow velocities, all measured in \nad\ absorption, can now be
compared over four orders of magnitude in SFR.

\begin{figure}
      {\includegraphics[scale=0.6,angle=-90,clip=true]{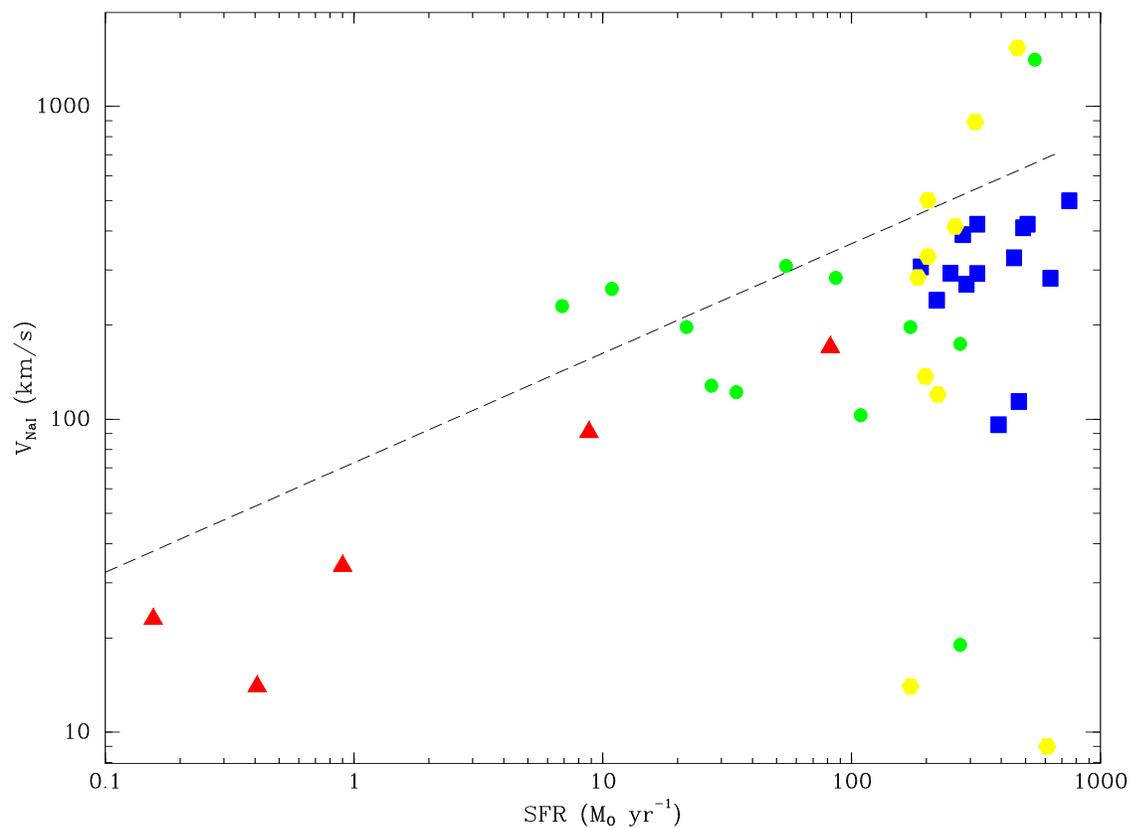}
	\hfill}
        \caption{Outflow velocities at line center vs galactic 
	  star formation rate.  The new data, 1~Jy ULIGs (Rupke \et
	  2002), LIGs (Heckman \et 2000), and dwarf starbursts (Schwartz \&
	  Martin 2004) are denoted by squares, hexagons, circles, and 
	  triangles, respectively.  These starburst systems represent
	  the highest surface brightness objects in their luminosity
	  class and are thought to define the maximum outflow velocities
	  reached. The dashed line represents the fitted upper envelope
	  when a simple model for projection effects is applied (see text
	  in \S 4.2 for detailsR).
	}
\label{fig:vsfr} \end{figure}

In \fig~\ref{fig:vsfr}, these velocities exhibit scatter at a given SFR, but
the upper envelope of the velocity distribution clearly rises with SFR.
The galaxies plotted in Figure~\ref{fig:vsfr} are all selected for their 
starburst activity.  They represent the highest areal star formation rate at 
a given luminosity and, presumably, the strongest outflows.  Velocity 
measurements in normal galaxies would  fill in the low velocity regions of 
Figure~\ref{fig:vsfr}.  The observed velocity spread may reflect 
mainly projection effects, as argued in \S4.3. In this case, 
a linear least squares fit to these data,
$ \log(V) = (0.35 \pm 0.06) \log (SFR) + 1.56\pm0.13 $
where the outflow velocity $V$ is in \kms\ and the SFR has
units \msunyr, should describe the slope of the upper envelope in 
\fig~\ref{fig:vsfr}. 

%The dashed line in \fig~\ref{fig:vsfr} shows this slope is a reasonable
%description of the upper envelope in all but the most luminous galaxies.

The dynamical masses of the galaxies are not as well constrained as the
star formation rates but provide a more fundamental quantity for comparing
the outflow velocities.   The three dwarf galaxies have rotation
speeds of 35\kms (\n1569; Stil \& Israel 2002 ), 
40\kms (\n4214; Walter \et 2001), and  87\kms 
(\n4449; Hunter, vanWoerden, \& Gallagher 1999) and are clearly less massive
than the LIGs and ULIGs. Ultraluminous activity, on the other hand, requires a 
large amount of gas, $ M_{gas}\sim 10^{10} $\msun, and a major merger
(Borne \et 2000). The expectation of massive progenitor galaxies is confirmed
by measured rotation curves and stellar velocity dispersions for three ULIGs 
in this sample (Genzel \et 2001).   In the
LIG luminosity range, however, the SFR is a poor surrogate for dynamical mass.  
The LIG subsample for which Heckman \et had rotation curves has a mean 
rotation speed of $199\pm45$\kms, which is less than that of the Galaxy
indicating some LIGs are not massive enough to become ULIGs. Some fraction
of LIG progenitors, however, are likely major mergers that were 
ultraluminous near perigalactic passage but faded during the long journey 
through apogalacticon.
This idea is supported by (1) the significant overlap in CO masses between
LIGs and ULIGs (Gao \& Solomon 1999), (2) the revived starburst activity 
suggested by the bimodel distribution of ULIG merger ages (Murphy \et 2001),
and (3) the wider mean separation of LIGs relative to ULIGs. Unfortunately,
most of the LIGs with measured rotation speeds are highly inclined disks
for which no outflow was detected by Heckman \et (2000). Consequently, 
rotation curves are only available for two of the LIGs shown 
in \fig~\ref{fig:vsfr}. 
Considering the paucity of data and poorly determined inclination corrections
for the outflow velocities, these results should be taken as preliminary. 
Nonetheless, the combined data do indicate a steep rise in outflow 
velocity with rotation speed, illustrated in Figure~\ref{fig:vterm_vc}.
This result is potentially of fundamental significance, so it will be
important to obtain more rotation curves.  For the galaxies without 
rotation curves, rough circular velocities estimated from line-widths
yield few velocities above than the preliminary relation, so
Figure~\ref{fig:vterm_vc} may actually provide a reasonable description of 
the upper envelope.

Absorption lines are ideal probes of  the kinematics of extended low 
density gas, so it is not surprising that outflows were one of the first features 
identified in spectra of high-redshift galaxies, whose optical spectra include
the plethora of strong absorption lines in the rest-frame ultraviolet bandpass. 
The choice of systemic velocity is critical
when comparing outflow speeds.  In particular, \lya emission is often 
seen redshifted relative to stellar features due to resonance scattering
off the receding side of the flow and therefore makes a poor zeropoint.
Among Lyman-Break-selected galaxies at $z \sim 3$,
the galaxy MS1512-cB58 has the highest quality spectra due its large lensing 
amplification.  The measured SFR, 40\msunyr, and outflow velocity,
255\kms (Pettini \et 2002) place it right
on the average relation for nearby LIGs.  The high-redshift objects
with bolometric luminosities in the ULIG range are selected in the
sub-mm (Ivison \et 2000).  
Rest-frame ultraviolet spectra show median offsets between the 
\Ha velocity and low-ionization ultraviolet resonance absorption lines of 
$\sim 650$\kms at luminosities of  a few hundred \msunyr (S. Chapman, pvt comm), 
which place them in the same part of \fig~\ref{fig:vsfr} as the local
ULIGs.  Outflows from high-redshift galaxies therefore appear to
obey the same scaling relations as winds from local starbursts.
 
Neutral outflows are not only faster in more massive galaxies, but
their terminal velocity also appears to increase almost linearly with the 
galactic rotation speed.  The bottom panel of \fig~\ref{fig:vterm_vc}
shows the terminal velocities are always 2 to 3 times the rotation speed.
This normalization is particularly interesting since the escape velocity  
is $\sqrt{2}$ (minimum) to 3.5 (isothermal halo extending to 100 kpc) times
larger than the circular velocity. The estimated escape velocities
are roughly 100\kms, 400\kms, and 900\kms in dwarf
starbursts, LIGs, and ULIGs respectively. The proximity of the upper
envelope in \fig~\ref{fig:vsfr} to these values provides important information
about the dynamics of the outflows and is discussed further in \S5.

\subsection{Velocity Variations from Sightline Orientation}

X-ray and optical imaging of nearby starburst galaxies indicate the outflow axis 
is aligned perpendicular to the disk plane. Identical bipolar outflows
viewed at random angles will yield a large range of measured \nad\ velocities,
so it is interesting to examine whether the fastest outflows  might
be found in systems with galactic disks oriented nearly face-on to
our sightline. This hypothesis is supported by the paucity of outflows in
the edge-on sample of LIGs (Heckman \et 2000). Inclination is only well defined 
for a few of the ULIGs due to both the lack of high-resolution imaging
and the on-going merger.
NICMOS images of  \i17208-0014 and \i10565+2448 show these disks
are oriented close to face on (Scoville \et 2000), and the outflow
velocities in both are among the largest measured at comparable SFR.
If large outflow velocities are purely an inclination effect, then
ULIGs like \i23365+3604 and \i18368+3549 must also be observed nearly face-on,
which is not inconsistent with their appearance in ground-based images. 
Large axial ratios indicate two ULIGs are likely viewed edge-on. 
Of these \i03521+0028 has very weak \nad, and \i11506+1331 has little net 
outflowing material in \nad.  Inclination therefore appears to 
have something to do with observed velocity spread at a given SFR. 

% the so $v_{max} = 1.85 v_{avg}$ 

\begin{figure}
      {\includegraphics[scale=0.9,angle=-90,clip=true]{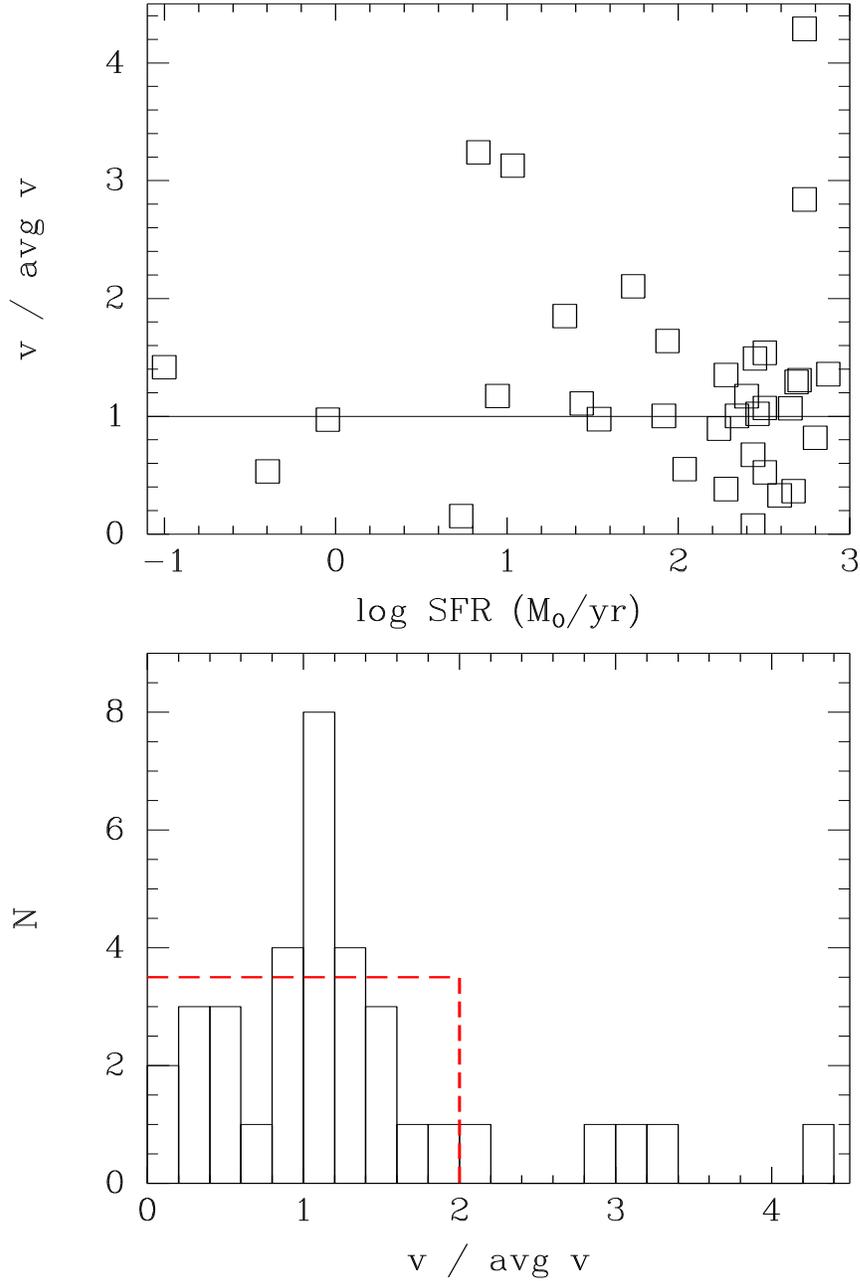}
	\hfill}
        \caption{(top) Distribution of outflow velocities normalized
	  by the mean at a given star formation rate. (bottom) Histogram 
	  of top panel for all values of the SFR.
	  The expected number of galaxies with
	  outflow velocity between $v_n$ and $v_m$ is 
	  $N(v_n,v_m) = N(v_{max},0) \frac{v_n - v_m}{2.0 <v>}$
	  illustrated in the bottom panel by a dashed line.  The four
	  objects with $v >> v_{max}$ present a significant departure
	  from the model.
	}
\label{fig:vh_vv} \end{figure}

\begin{figure}
      {\includegraphics[scale=0.9,angle=-90,clip=true]{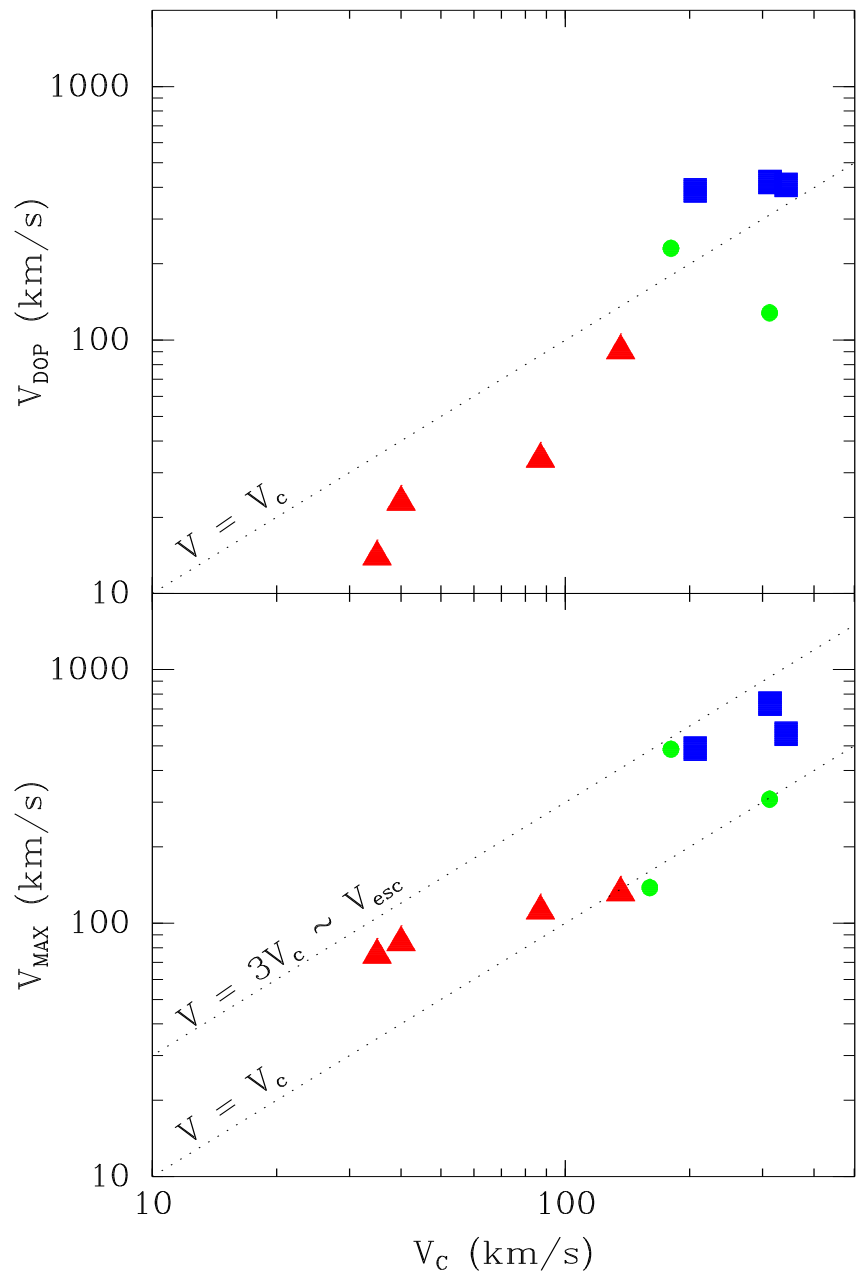}
	\hfill}
        \caption{{\it (top)} Outflow velocity at \nad\ line center versus 
	  galactic rotation speed. {\it (bottom)} Terminal \nad\ velocity  
	  versus galactic rotation speed. 
	}
\label{fig:vterm_vc} \end{figure}

To examine inclination effects quantitatively, consider 
a simple  model where the flow is perpendicular to the disk plane. 
The average polar viewing angle is $i = 57\deg$, and the average projected
velocity is $0.5 v_{max}$. It follows that, at any particular SFR, the true
outflow velocity is twice as  large as the average projected outflow velocity
measured.
Multiplying the fitted $v_B ~vs~ SFR$ relation by 2.0 yields the dashed line 
sketched in \fig~\ref{fig:vsfr}. This estimate of the de-projected outflow
velocity describes the upper envelope adequately 
for all but a few, particularly luminous, objects. 
The distribution of observed velocities
about the mean is shown in Figure~\ref{fig:vh_vv}, where the data were 
divided by the mean velocity at the appropriate SFR and then binned according
to their deviation from the mean. The naive, planar flow model predicts
a constant number of galaxies with increasing $v/v_{avg}$ up to $v_{max}
= 2.0 v_{avg}$. The data show more galaxies with outflow speeds near the 
average than predicted, but this may be consistent with a more realistic 
representation of the wind geometry where the velocity field has a radial
component. Projection effects, however, fail to explain the high velocity 
tail in the distribution, and this shortcoming is further discussed in \S5.

\subsection{Merger Sequence}

In addition to inclination, the other property that clearly varies
among the ULIGs is the dynamical age of the merger, which is constrained by 
morphology (Murphy \et 2001). Visual and near infrared imaging of ULIGs 
drawn from the 2~Jy sample show galaxy pairs, tidal tails, and double nuclei 
(Murphy \et 1996). Inspection of these images allows some ULIGs to
be identified as first passage, late passage, and fully merged. The ESI 
spectra resolve a number  of nuclei, previously classified as 
single nuclei, into double nuclei. The outflow velocities do not 
correlate with dynamical age, but the strength of the wind does show
some interesting trends that may provide some insight about the role of 
the winds in the ULIG-QSO transition.

Highly eccentric orbits are required to create an intense, short-lived 
disturbance which centrally concentrates the interstellar gas, so  
the galaxies pass rapidly on their first encounter. Short tidal features 
are one signature of first passage because it takes time for tidal features 
to grow. Disk-like morphology also indicates not enough time has elapsed for 
the stellar orbits to reflect the transferred orbital energy. 
Of the 18 ULIGs discussed in this paper, five are plausibly on
their first passage. Short tidal features are seen in \i10565+2448, 
\i11506+1331, \i19458+0944, and \i17208-0014, and the debris in
\i00153+5454 is confined to the region around the distorted disks.
These are all double nuclei systems.  Of the five strongest \nad\ lines 
(i.e. equivalent width $> 7$\angs), two are found among these 5 galaxies.
The other three strong lined systems are caught during the second pass.
In \fig~\ref{fig:merg_seq}, these two are pictured along with
their line profiles to represent the first-pass mergers.

The duty cycle of the ultraluminous phase must be $\sles\ 20\%$
because the estimated dynamical timescales for the mergers,
$5 \times 10^8$~yr to $15 \times 10^8$~yr (Murphy \et 1996),
exceed the gas consumption timescales, $\sim 10^8$~yr. Most of the
time between passages is spent near apogalacticon where the galaxies
are unlikely to be ultraluminous. None of the systems with
strong winds are found in a widely separated phase. The lack of tails and 
largely symmetric structure in \i00188-0856, \i16487+5447, and \i19297-0406 
point toward a stage just before the second encounter. The relatively large 
separations 20.2~kpc, 5.3~kpc, and 4.4~kpc are consistent with this picture.  
Their \nad\ absorption, \fig~\ref{fig:merg_seq}, is quite weak,  
comparable to \nad\ line strengths in LIGs.  These observations indicate
the wind strength decreases when the galaxies are far apart between 
encounters. 
Three systems in the ULIG subsample can be identified with the second (or nth) 
passage by their double nuclei, large tidal structures, and/or more
spheroid-like than disk-like structure.  The identification as 2nd-passage
systems is most secure for \i15245+1019, \i20087-0308, and \i18368+3549, 
which have particularly long tidal tails.  All three of these ULIGS present
very large \nad\ equivalent widths.

The nuclei were not resolved in 7 systems, and the interstellar \nad\ lines 
are generally weaker in these single nuclei systems. The most recently 
merged systems still present large, diffuse tidal structures as seen in 
\i00262+4251, \i03521+0028, \i16090-0139, and \i23365+3604. The line
strengths are moderately strong falling between those of systems on the 
nth-passage and the most advanced mergers in this sample.  The other
two systems, \i03158+4227 and \i08030+5243, present quite symmetric
and much fainter tidal debris; and the interstellar \nad\ absorption is
weak. These results suggesting the wind is dying down in the older systems.

%No image was available for \i10494+4424.

The combined small size of this ULIG sample and the large variations
in the accuracy of the dynamical ages estimates makes a robust statistical
analysis unfeasible.  However, this preliminary exploration strongly
suggests that the strongest winds are found in systems near perigalactic 
passage.  Furthermore, the winds appear to weaken when
(1) the galaxies are far apart and (2) the merger is complete.
The starburst probably dies down due to  growing scarcity of gas.
Further investigation may determine whether the gas is simply consumed,
removed by the starburst, or removed by a nascent AGN-driven outflow.

\section{Acceleration of the Cool Wind}

Consistency between the \x and \nad\ measurements is an important test 
of the standard dynamical model for starburst winds, which assumes a 
thermally-driven hot wind accelerates cooler, entrained gas
(Chevalier \& Clegg 1985; Heckman, Armus, \& Miley 1990). 
Numerical simulations based on this idea reveal that large vortices
develop at the shear layer between the hot, bipolar wind and the gaseous
galactic disk (e.g. see Figure 11 in Heckman \et 2000). The \nad\ absorption
is thought to be a good tracer of the cool gas entrained along these
interfaces.  The motion of the clouds is therefore expected to reflect
the properties of the hot wind fluid.
The X-ray surface brightness is biased towards the densest part of 
the wind, however; and observations may not detect the hottest component of an
outflow. Nonetheless, two factors motivate a simple model where the
wind temperature does not vary much with starburst luminosity or mass.
First, the \X temperature of the mass-carrying component of the hot wind
does not show a significant dependence on galactic mass (Martin 1999;
Heckman \et 2000). Second, theoretically, the hot wind temperature 
is simply the amount of thermalized supernova energy per unit mass of
entrained material. The energy and entrained mass may both scale linearly
with starburst luminosity producing similar temperatures in all starburst
winds. Another model, especially applicable to the ULIGs, 
is that radiation pressure on dust grains in the clouds accelerates
the cool wind.  Similar ideas regarding this radiative feedback have
been advanced for limiting the mass of OB associations (Scoville \et 2001).
This idea is particularly interesting since
the associated Eddington-like luminosity for galaxies would limit
the maximum brightness of a galaxy in a given gravitational potential
(Scoville 2003; Murray, Quataert, \& Thompson 2004). In what
follows I give some initial interpretation of the data presented in
this paper.  A more extensive discussion is being developed in a 
series of papers with Eliot Quataert, Todd Thompson, \&
Norm Murray.

\subsection{Supernova-Driven Winds}

The rise in \nad\ outflow velocity with galactic mass, Figures~\ref{fig:vsfr} 
and \ref{fig:vterm_vc},  appears to be in direct conflict with the
relatively uniform \x temperatures measured over a similar mass range
(Martin 1999; Heckman \et 2000; Martin 2004). Gas at
$T_x \sim 8.9 \times 10^6$~K (0.76 keV) in a hot bubble allowed to stream 
through a ruptured supershell accelerates to a terminal 
velocity of $\sqrt{3} c_s \approx 500 (kT_x/0.76 {\rm ~keV})^{1/2} $\kms 
in the intergalactic medium. This velocity is remarkably
similar to the speed of the \nad\ clouds in the ultraluminous starbursts.
The mystery is thus reduced to understanding  why the cool gas is {\it not}
accelerated to the wind velocity in less luminous starbursts. 

One can imagine several reasons why the clouds might not reach the wind 
velocity.  It would be interesting, for example, to examine whether the
gaseous halos of dwarf galaxies can prevent the 
formation of a free-flowing wind (cf. Silich \& Tenorio-Tagle 2001;
Mac Low \& Ferrara 1999).  The \x temperatures indicate that winds 
in massive galaxies and dwarf galaxies start out with similar specific thermal 
energy, but the observations do not have enough spectral resolution to directly 
measure the kinematics of the hot fluid.\footnote{An even hotter component may
  be present in massive galaxies, but current limits on its emission 
  measure suggest it carries a tiny fraction of the mass.  Indeed, 
  as long as more powerful winds entrain more interstellar gas, it is
  hard to avoid similar mass-weighted \x temperatures among starbursts
  of all luminosities.}
Alternatively, the clouds might be photoionized or disrupted via 
hydrodynamic instabilities before they reach the velocity of the hot wind,
but this solution is not appealing, however, since it requires a complicated
scaling of cloud lifetime with starburst luminosity.
A simpler hypothesis is that the hot wind in lower luminosity
starbursts does not carry enough momentum to accelerate clouds to
the hot wind velocity. If correct, then the cloud velocities should
tend toward a constant luminosity in more luminous starbursts.

% AGHH: IT'S THE DENSITY OF CLOUDS THAT IS HIGH NOT THE DENSITY OF GAS THAT
% MAKES UP THE CLOUDS.
%\footnote{The wind is 
%  expected to accelerate over a shorter 
%  distance in the dwarf starbursts owing to the shallow gravitational 
%  potential. Mass conservation requires a steeper density gradient in
%  the cloud population,  and therefore a steeper pressure gradient if
%  the cloud temperature is regulated.}

The upper envelope of the velocity -- SFR relation in Figure~\ref{fig:vsfr}
may flatten beyond a SFR of $\sgreat\ 10$\msunyr. The data points 
in this region are sparse, but most would agree that a break in the slope is 
not ruled out. Since \x observations suggest
the mass flux in the hot wind is similar to the SFR, or equivalently, that 
the mass loading is about a factor of ten larger than the mass ejected 
by supernovae and stellar winds (Martin 1999; Martin \et 2002), the rate of
momentum injection into the hot wind is relatively well constrained. A given 
column of cool gas will be accelerated to a maximum velocity determined
by the equation of motion.  Provided the cloud lifetime exceeds the outflow 
timescale, the ram pressure of the hot flow accelerates a cloud of mass $M_c$ 
and area $A_c$ as
\begin{equation}
M_{c} \frac{dv_{c}}{dt} =  \dot{m}_{w}(v_w - v_{c})\frac{A_c}{\Omega r^2}
 -  \frac{G M(r) M_{c}}{r^{2}}, 
\label{eqn:f=ma} \end{equation}
where $v_w$ is the velocity of the hot wind streaming out through solid 
angle $\Omega$.  The momentum injection rate is $\dot{p}_{w} = 
\dot{m}_w v_w = \Omega \rho_w r^2 v_w^2$, which is about 
$5 \times 10^{33} (SFR/1\msunyr)$~dyne in a starburst 
(Leitherer \et 1999).\footnote{The Starburst~99 population synthesis code was 
used to estimate the rate at which supernovae and stellar winds supply
momentum, i.e. $\dot{p}_* = \dot{m}_{*,ej} v_{*,ej} = 
\sqrt{2\dot{m}_{*,ej} L_w}$. A Salpeter stellar 
  mass function from 0.1\msun\ to 100\msun\ was adopted to be consistent 
  with \fig~\ref{fig:vsfr}, and the star formation rate was assumed to 
  be constant for 100~Myr.  At an age of 10~Myr, $\dot{p}_{w}$ would be about 
  a factor of two lower.}
The absorbing clouds must enter the hot wind at relatively low velocities
because the absorption line profiles extend smoothly to the systemic velocity.
Replacing the time derivative in Equation~\ref{eqn:f=ma} with $v dv/dr$,
ignoring the gravitational deceleration, and integrating outward from the 
sound speed of the cool clouds, i.e. $\sim 6$ km/s, 
yields  terminal cloud velocities that increase with SFR as illustrated
in  \fig~\ref{fig:v14_sfr}.  Up until the point where the clouds are moving
with the hot wind, these solutions are well approximated by
taking $v_w - v_c \approx v_w$ in Equation~\ref{eqn:f=ma} giving
terminal velocity
\begin{equation}
v_{term} = \left( \frac{3 \dot{p}_{w}}{2 \Omega \rho_{c} R_{0}^2}
\right)^{1/2}
 = \left( \frac{3 \rho_w(R_{0})}{2\rho_{c}} \right) ^{1/2} v_w
\end{equation}
for clouds of density $\rho_{c}$ launched from
radius $R_0$ (e.g. Murray \et 2004, Appendix A).   Substituting
$N_c = \rho_c  \mu^{-1} m_p^{-1} R_c$, the fiducial wind speed of 500\kms,
$\dot{m}_w = \lambda \dot{M}_*$,
and an arbitrary launch radius of 200~pc, the maximum cloud velocity can
be written
\begin{eqnarray}
\frac{v_{term}}{v_w} = 
\left( \frac{\lambda\dot{M}_*}{0.24 ~{\rm M_\odot~yr}^{-1}}
                      \right)^{0.5} \times \nonumber \\
		       \left( \frac{10^{20} {\rm cm~}^{-2}}
			                          {N_H} 
			 \frac{200 ~{\rm pc}}{R_0} \frac{500 
			   {\rm ~km~s}^{-1}}{v_w}
			 \frac{\pi}{\Omega}
			  \right)^{0.5},
\end{eqnarray}
where $\lambda \sim 1$ and $\mu$ was taken to be 1.4.
This approximation indicates the critical SFR required to accelerate a cool 
gas column of $N_H \sim 10^{20}, 10^{21}, {\rm ~and~}, 10^{22}$~cm$^{-2}$ to 
500\kms\ is 0.24\msunyr, 2.4\msunyr, and 24\msunyr, respectively. Taking
proper account of the relative cloud -- wind velocity in \fig~\ref{fig:v14_sfr}
shows somewhat  higher SFR's, $\dot{M}_* \sim 0.7, 7, 70$\msunyr, are
required to actually flow with the wind. Inspection of \fig~\ref{fig:v14_sfr}
indicates that any combination of launch radius and column density
satisfying $(\mu N/10^{22} {\rm ~cm}^{-2}) (R_0/200 {\rm ~pc}) \sim 1$
is a good description of the upper envelope of the outflow velocities
for all but a few extremely luminous ULIGs. 
The implied rate of momentum injection,
$\dot{p_w} = 2/3 \Omega  \mu m_p N_H R_0 v_{\infty}^2$ or
\begin{equation}
\dot{p_w} = 8.0 \times 10^{34} {\rm ~dyne~} (\Omega/\pi)
N_{22} R_{0,200} v^2_{\infty,500},
\end{equation}
which, for any solid angle $\Omega$, is easily supplied 
by a SFR of 70\msunyr\ since $\dot{p}_* = 3.5 \times 10^{35}$~dyne.
Supernova-driven winds can,
it turns out, describe the observed increase in \nad\ velocities from 
${\rm SFR} ~ 0.1$ to 
roughly 100\msunyr\ and still be consistent with \x temperatures that are 
independent of galaxy mass.

It should be emphasized that the terminal cloud velocity is sensitive to the 
launch radius because most of the cloud acceleration happens at small 
radii where the ram pressure is largest. 
If the size of the starburst region, and consequently the launch region
for the wind,  grows as the SFR increases, then the terminal velocities
rise more slowly than $SFR^{0.5}$. Figure~\ref{fig:v14_sfr} illustrates
the constant surface brightness case previously suggested by 
Heckman \et (2000), which implies $v_{term} \propto {\rm ~SFR}^{1/4}$. 
This scaling for the size of the launch region fails to reproduce
the steep slope of the outflow velocity envelope from dwarf starburst to LIG
luminosities.

It is not obvious that the hot wind would naturally tune the terminal cloud 
velocities to match the rotation speed of the parent galaxy as suggested 
by \fig~\ref{fig:vterm_vc}.  This correlation implicitly requires 
$\dot{p}_w$ to be connected to the halo mass, presumably via the SFR.
Some simple recipes are explored here.  Using an isothermal halo with 
velocity dispersion $\sigma$ to model the halo mass distribution,
$M(r)  = 2 G^{-1} \sigma^2 r$,  Equation~\ref{eqn:f=ma} becomes
\begin{equation}
\frac{dv_c}{dr} = \frac{-2\sigma^2}{v_c r} + \frac{3\dot{p}_w}{4 N_H \mu m_h
\Omega} \left(1 - \frac{v_c}{v_w} \right) \frac{1}{v_c r^2},
\label{eqn:grav} \end{equation}
which is easily solved for $v(r)$ when $v_c << v_w$.  
In order for the clouds to move out a significant distance, the solution
suggests the characteristic flow velocity $v_{char} = 
\sqrt{3\dot{m_w}v_w / (2\Omega N_H R_0 m_p \mu)}$
must exceed $2\sigma$ (e.g. Murray, Quataert, \& Thompson 2004). 
For purposes of illustration, suppose the star formation rate scales as 
$SFR \propto f_g M_{tot} / t_{dyn} \propto \sigma^3$, where $f_g M_{tot}$
is the gas mass of the galaxy.  This scaling is
plausible since the mass of a galaxy increases as $\sigma^3$ and the 
dynamical timescale of a disk, $R_d / V$, is independent of mass (Mo, Mao, 
White 1998). Normalizing to the stellar velocity 
dispersions in \i23365+3604, \i20087-0308, and \i17208-0014 (Genzel \et 2001) 
gives  $SFR \approx 500 \msunyr (\sigma / 210 \kms)^3$. Using this relation
and integrating Equation~\ref{eqn:grav} from low initial velocity at
$R_0 = 200$~pc,  clouds accelerate to the velocities shown in the top
panel of Figure~\ref{fig:vterm_model} before gravity slows them down.
Choosing $(N_{H}/ 10^{22} $~cm$^{-2})(R_0 / 200 {\rm ~pc}) \approx 1$ again
provides the best normalization to the data. With this $SFR \propto 
\sigma^3$ recipe, the cloud velocities  rise steeply with the rotation speed 
because star formation is greatly suppressed in the low mass galaxies.
Scaling the SFR as  $\sigma^2$ or $\sigma$ produces a much more gradual
rise in outflow speed with rotation speed, and these models are not
consistent with the data.

% , and it follows that the maxmimum column the wind can move is 
% $N_H (R_0/200 {\rm ~pc}) \sles\ 4.6 \times 10^{20}
% {\rm ~cm}^{-2} 4\pi / \Omega (\dot{M_*} / 1 \msunyr)$  

% The cloud column density here is $N_H \mu m_p = \rho{c} R_0$ where $\mu$ is 
% the mean atomic mass per H atom.

This exercise demonstrates that ram pressure from a supernova-heated 
wind accelerates cool gas to a maximum velocity determined by the 
temperature of the momentum-carrying component of the hot wind. Outflow
velocities measured for the 2~Jy sample of ULIGs are consistent with a 
maximum wind speed near $v ~500$\kms, but more measurements will be 
required to confirm or refute this suggestion. Indeed,  in 
\fig~\ref{fig:v14_sfr},  four galaxies from the Rupke \et and Heckman 
\et samples present outflows significantly faster than this fiducial
maximum.  Several of the 2~Jy ULIGs also show absorption, albeit far from 
line center, at velocities in excess of 500\kms.
Failure of the outflow measurements
to define an asymptotic velocity might reveal (1) the presence of a hotter 
wind component or (2) an acceleration mechanism other than supernova.
The awkwardness of the supernova-driven wind model for explaining the 
terminal velocity -- rotation speed correlation motivates some 
consideration of alternative acceleration mechanisms.

%The case for a hotter component of the wind has been discussed by 
%Strickland \& Stevens (2000) and Strickland \et (2002). Although  hotter,
%undetected gas is almost certainly present, this phase contains little
%mass and does not necessarily dominant the momentum flux.

% the size of the disk grows with the virial radius, which is proportional to 
% $\sigma$ leaving 

% total mass of the galaxy, $M = 2 G^{-1} \sigma^2 r_{200}$,
% grows with cosmic time as $\sigma / H(z)$ 

% The adopted starburst age has little impact on the normalization 
% of $v_{term}$ if the starburst size is allowed to increase as
% $R_{SB}150 {\rm ~pc~} \left( \frac{SFR}{1 {\rm M_\odot ~yr^{-1}}} 
% \right) $	(Meurer \et 1997) since  both  and $R_{SB}^2$ 
 
% the models set $SFR = \dot{M_w}$ 

% The normalization is well-constrained except for average cloud density,
% $v_{term} = 930 \kms \sqrt{\pi \Omega^{-1} (1.67 \times 10^{-24} 
% {\rm ~g cm}^{-3} \rho_{cl}^{-1}}) $

%The actual velocity attained depends on the cloud's intial launch radius and
%density as well as the amount of momentum supplied by stars in a halo of a 
%given mass. 

% The clouds begin to 
%decelerate at  a radius
%\begin{equation}
%r_{crit}^2 = \frac{\dot{p}_{w}}{\Omega \sigma^2 \rho_{cl}},
%\end{equation}
%or
%\begin{equation}
%r_{crit} \approx  1.0~{\rm ~kpc~} \frac{\pi}{\Omega} \left( \frac{\dot{M}}{1 \msunyr}
%\right)^{1/2} 
%\left( \frac{1.67 \times 10^{-24}}{\rho_{cl}} \right)^{1/2} \frac{100 {\rm ~km/s}}{\sigma}.
%\end{equation}

\subsection{Radiatively-Driven Winds}

Stellar wind velocities exhibit a scaling similar to that found here
for galactic winds.  The terminal wind 
velocities from central stars of planetary nebulae are typically
4.4 times the escape velocity at the stellar surface (Lamers \&
Cassinelli 1999).  Is this remarkable coincidence an indication
that the underlying wind physics is similar?  The correlation seems
natural in stellar winds. The wind is accelerated by radiation
pressure on dust grains, and the maximum surface luminosity is limited 
by the requirement of hydrostatic equilibrium.  In galaxies,
the SFR determines the momentum of the hot wind, but the
relation between galaxy mass and SFR is complicated by 
variable  quantities such as the gas fraction and the star formation
efficiency. Radiatively-driven winds on a galactic scale have an
appealing property, namely that the galactic luminosity cannot increase 
much beyond an effective Eddington-like luminosity.

The radiative force exerted on the cool gas from dust absorption 
usually dominates over that from photoelectric absorption by gas.\footnote{
  For this to be strictly true, the ionization parameter must exceed 
  $U_0 \approx 0.01$ (Dopita \et 2002).
  This is easily seen by equating the
                  hydrogen recombination rate and the rate of
		  dust absorption per H atom.
		  $ \frac{S}{nc} > \frac{\alpha_B}{\kappa c} 
		  \frac{\bar{\epsilon_{FUV}}}{\bar{\epsilon_{UV}}}$
		  where $S$ is the photon flux.}
For the radiatively-driven wind concept to be viable for galaxies,
the dust grains need to survive in the outflow {\it and} pull the gas along.
Dust grains in the interstellar medium are charged and typically
couple electrostatically to the gas (Draine 2003). 
Sputtering following the passage of a shock 
destroys grains, but calculations suggest significant amounts of dust survive  
in radio galaxies (De Young 1998; Villar-Martin \et 2001).
Several types of observations strongly suggest that starburst winds are
indeed dusty.  First, the \nad\ line is a resonance line, so the absorbed 
photons are re-emitted in random directions. The absence of \nad\ 
{\it emission} in the ULIG spectra presented in this paper suggests the
re-emitted photons are absorbed by dust (and converted to heat)
before they can escape the galaxy. Second, the
thermal emission from cold dust in starburst galaxies has a significantly 
larger scale-height than the stars indicating dust is lofted above the region 
where it forms (Engelbract \et 2004, NGC 55; Popescu \et 2004, NGC 891). 
Third, extended red emission, a broad emission band seen in many dusty 
astrophysical objects, has been detected in the halo of the superwind galaxy 
M82  (Perrin, Darbon, Sivan 1995; Gordon, Witt, \& Friedmann 1998). 
Furthermore, recent GALEX imaging of M82 show ultraviolet halo emission
significantly brighter than expected from the nebular continuum, which
likely indicates the light is scattered by small grains (Chris Martin, pvt 
comm).  The dynamical implications of dust in these winds needs to be
considered.

% ERE ==> contains C

% Dopita \et (2002) has argued that dust will survive in the extended 
% NLR of AGN. 
 
% (Howk ----; ; Zaritsky ??).

% $\kappa \sim 10^{-21}$~cm$^2$ per H atom.

%\begin{equation}
%\frac{L}{c} \sgreat\ \frac{GM(r)M_c(r)}{r^2},
%\end{equation}

Little ultraviolet radiation emerges from ULIGs implying these
galaxies are optically thick to much of the intrinsic starburst
luminosity. The radiative force on the gas clouds is $F_{rad} = 
M_c c^{-1} \int \kappa_{\nu} F_{\nu} d\nu$, where
the absorption opacity of the dust and gas mixture, $\kappa$, drops from 
600~cm$^2~{\rm g}^{-1}$ near 1000\angs\ to 
100~cm$^2~{\rm g}^{-1}$ in the U~band (Li \& Draine 2001).
In the optically thick limit, the radiation force reaches $L/c$
and balances the gravitational force on the cool interstellar gas
when the luminosity
reaches $L_{crit,ED} = 4G^{-1} c f_g\sigma^4$.  For a fiducial
halo velocity dispersion of 150\kms, the critical luminosity required 
to radiatively accelerate most of the cool interstellar gas is then
\begin{equation}
L_{crit,ED} = 2.4 \times 10^{12}{~\rm L_\odot} \left( \frac{\sigma}
{150 {\rm ~km/s~}} \right)^4 \left( \frac{f_g}{0.1} \right).
\end{equation}
The fraction of the  galactic mass in the cool wind, $f_g$, 
will be less than 0.1, so the ULIG luminosities appear high enough 
to radiatively accelerate significant amounts of cool gas and dust.
Assuming a starburst duration of 100~Myr (Leitherer 1999) and a Salpeter 
($\alpha = 2.35$) initial mass function from 0.1 to 100\msun,
the corresponding star formation rate where radiation pressure becomes 
important for wind acceleration is
\begin{equation}
\dot{SFR}_{crit} \sim
400 \msunyr\ \left( \frac{\sigma}{150 {\rm ~km~s}^{-1}}
\right)^4 \frac{f_g}{0.1}.
\end{equation}

Any additional (besides supernova) acceleration mechanism should
predict the highest outflow speeds and/or explain why most outflows are
tuned to the rotation (or escape) speed.  The latter is clear.
Once the starburst luminosity exceeds $L_{crit,ED}$, the cloud is 
accelerated from rest at $R_0$ to $v(r) = 2\sigma \sqrt{(L/L_{crit,ED} - 1) 
ln(r/R_0)} $ (e.g. Murray, Quartaert, and Thompson 2004).  The weak
radial dependence indicates the terminal cloud velocity depends
mainly on the ratio $L/L_{crit,ED}$. Figure~\ref{fig:vterm_model} (bottom
panel) demonstrates that $L/L_{crit,ED} \sim 2$ provides an excellent
description of the correlation between outflow velocities and rotation speed.  

The outflow speeds exceeding 500\kms\ could have several explanations.
If radiation pressure is the dominant acceleration mechanism, then
starbursts are not the source of luminosity because extrapolation of
the velocities in \fig~\ref{fig:vterm_model} to 1500\kms\ implies rotation
speeds larger than galaxies.  AGN would have to supply the radiative
acceleration. Ram-pressure could also accelerate cool gas to these high 
velocities if the clouds are launched from small radii associated with
AGN rather than starburst radii.
Inspection of the starburst spectra with outflows faster than 500\kms 
indicates IRAS-FSC05024-1941 (1538\kms, Rupke \et 2002) 
has cool infrared colors but is optically identified as a Seyfert~2;
IRAS11119+3257 (1410 \kms, Heckman \et 2000) has warm infrared color.
It therefore seems quite likely that both these objects have active nuclei,
and one can speculate that AGN are responsible for the abnormally high
velocities.  A much larger database will clearly be required to determine
the relative roles of AGN and starbursts in driving winds. Among the
the 2~Jy~CO-subsample, however, only \i11506+1331 shows Seyfert~2 activity,
and it does not present an unusually fast wind.

\section{Conclusions} 

Doppler shifts measured from interstellar absorption lines in starburst spectra
reveal cool gasesous outflows.  The fraction of starbursts presenting outflows
increases with starburst luminosity reflecting either the brief duty cycle
of the ultraluminous phase or the large solid angle subtended by these winds.
The cool outflow does not entirely cover the optical nucleus indicating
clouds or shells are a better description than a continuous fluid. The covering 
fraction approaches unity in the galaxies with the strongest winds (i.e. largest 
equivalent width), and these merging systems are usually found to be near perigalactic 
passage. More widely separated systems and single-nuclei systems generally have
lower covering factors and line strengths.  These results lead to the
conclusions that (1) much of the mass loading occurs very near perigalactic
passage and (2) the mass loading is gradually shut down as the nuclei
finish merging.  The mass entrainment rate of cold clouds is quite large, 
possibly comparable  to the SFR, and is discussed in more detail in Paper II.

Galaxies with higher SFRs accelerate the absorbing clouds to higher velocities.
Projection effects introduce a large range in measured velocities at a given SFR,  
but the maximum outflow velocity rises roughly as $SFR^{1/3}$. 
This trend challenges the hypothesis that the clouds are accelerated by the hot winds,
whose emission-measure weighted temperatures appear to vary little with circular
velocity.  The simplest explanation is that the clouds in the ultraluminous 
starbursts reach terminal velocities comparable to the velocity of the hot winds, 
and that the low \nad\ wind velocities in dwarf starbursts reflect a
shortage of momentum in dwarf starburst winds.

However, in order for momentum-driven winds to explain the intriguing similarity 
between the  terminal cloud velocity and the galactic escape velocity, the
rate of momentum injection must be tightly correlated with the depth of the
gravitational potential.  Radiatively-driven winds naturally provide this
connection via a galactic equivalent to the stellar Eddington luminosity, while
rather ad hoc recipes between the SFR and dark halo mass must be cooked up
for supernova-driven winds.  A highly simplified description of how dusty
clouds would be accelerated by the starburst continuum luminosity demonstrates
that ULIG luminosities are sufficient to accelerate significant columns of
cool gas to the observed \nad\ velocities.

These results draw attention to the importance of feedback in very massive galaxies.
Preliminary estimates of the mass loss rates suggest a significant fraction of the ISM 
participates in the cool outflows. This result is surprising in the context of
energy-driven winds because a large fraction of the supernova energy is expected to be 
radiated away in the dense molecular ISM of an ULIG.  Much as supernova feedback is most effective
in dwarf galaxies where hot gas can escape from the halo, radiative feedback is naturally
strongest in the densest, most luminous galaxies.  Hence, radiative feedback may 
help explain the differences between the shape of the halo mass function and the 
galaxy luminosity function at the high-mass end. 
  The dynamical properties of the stars in ULIGs do appear to be 
consistent with those of field ellipticals (Genzel \et 2001), so it is also
appealing to associate the winds seen in ULIGs with the event that shuts the 
star formation down quickly enough to create high ratios of alpha elements to 
iron in elliptical galaxies.  

Filling in \fig~\ref{fig:v14_sfr} and \fig~\ref{fig:vterm_model} with more observations 
will help clarify some of these isssues. Also, since radiatively-driven winds do not 
require a hot phase,  X-ray observations of these ULIGs may distinguish 
acceleration mechanisms. Spitzer and GALEX observations will be instrumental for constraining
the amount of extraplanar dust in starbursts and its composition.
A few of the most luminous objects present velocities too high
to explain with starburst winds, so it will be possible to explore the nature of outflows 
in warm ULIGs during the SB/AGN transition. More detailed dynamical modeling
of the survivability and fate of the cool clouds as well as a more realistic
treatment of  radiation pressure will allow a more detailed comparison to the
data (e.g. line widths, spatial gradients, etc.) and thereby a more
stringent test of the ideas presented here.

\acknowledgements{ 
The observations presented here would not have been possible without
the R band images of the galaxies provided by Tom Murphy. The author thanks 
Lee Armus for help with some of the spectral fitting and enthusiasm for 
starting this project. Discussions with Norm Murray, Todd Thompson, and 
Eliot Quataert regarding the theoretical interpretation of the data 
presented here are greatfully acknowledged.  This paper was completed at the 
Aspen Center for Physics where discussions with Karl Gordon and Timothy 
Heckman improved the content of this paper. Funding for this project was 
provided by the David and Lucile Packard Foundation and the Alfred P. Sloan 
Foundation.  Finally, the author thanks those of Hawaiian ancestry on whose 
sacred mountain these data were obtained.
}

% Facilities: \facility{Keck(ESI)}

%% TABLES
\clearpage

\begin{table}
\caption{Sample of Galaxies}
\begin{tabular}{lllll}
\hline
 Object &	  z &       $L_{FIR}$ &  $L_{IR}$ & SFR \\
        &        (CO)   &  (\lsun)  & (\lsun) & (\msunyr) \\
\hline
\hline
IRAS00153+5454 & 0.1120  &  11.94 &   12.27 & 320 \\ 
IRAS00188-0856 & 0.1285	 &  12.18 &   12.41 & 450 \\	
IRAS00262+4251 & 0.09724 &  11.90 &   12.16 & 250 \\	
IRAS03158+4227 & 0.1344	 &  12.39 &   12.64 & 750 \\	
IRAS03521+0028 & 0.1519	 &  12.33 &   12.57 & 640 \\	
IRAS08030+5243 & 0.08350 &  11.82 &   12.05 & 190 \\	
IRAS10494+4424 & 0.09231 &  11.99 &   12.23 & 290 \\	
IRAS10565+2448 & 0.04311 &  11.81 &   12.05 & 190 \\	
IRAS11506+1331 & 0.1273	 &  12.11 &   12.35 & 390 \\	
IRAS15245+1019 & 0.0757  &  11.81 &   12.10 & 220 \\
IRAS16090-0139 & 0.1336	 &  12.34 &   12.57 & 630 \\	
IRAS16487+5447 & 0.1036  &  11.98 &   12.29 & 330 \\
IRAS17208-0014 & 0.04282 &  12.23 &   12.45 & 490 \\
IRAS18368+3549 & 0.1162	 &  12.03 &   12.27 & 320 \\	
IRAS19297-0406 & 0.08573 &  12.21 &   12.44 & 470 \\	
IRAS19458+0944 & 0.1000	 &  12.15 &   12.39 & 420 \\	
IRAS20087-0308 & 0.1057	 &  12.23 &   12.47 & 510 \\	
IRAS23365+3604 & 0.06448 &  11.96 &   12.20 & 280 \\	
\hline
\end{tabular}  

Table Notes:
     (1) Object name.
     (2) Redshifts from Solomon et al. (1997) except
     IRAS~00153+5454, IRAS~15245+1019, and IRAS~16487+5447
     which are from Dr. Aaron Evans (pvt. comm.).
     Velocities have errors of $\pm cz \approx 20$\kms.
     (3) The far-infrared luminosity computed from 
     $L_{FIR} = 3.86 \times 10^5 \lsun d_L^2 (2.58 F_{\nu}(60um) + f_{\nu} (100um))$,
     where the flux density is in Janskys and the 
     luminosity distance (h=0.7, $\Omega_0 = 0.3$ and $\Omega_{\Lambda} = 0.7$)
     is in Mpc.
     (4) The estimated 8 - 1000 um luminosity $L_{\rm IR}$
     (h=0.7, $\Omega_0 = 0.3$ and $\Omega_{\Lambda} = 0.7$).
     The IRAS 12um fluxes
     for the galaxies in this sample are upper limits, so the median
     flux ratios of bright ULIGs with IRAS detections in all four
     bands were used to estimate $L_{\rm IR}$ as described in  
     Murphy \et (1996).
     (5) The star formation rate estimated from $SFR = L_{\rm IR} / 5.8 
     \times 10^9$ 
     \lsun (Kennicutt 1989), where the luminosity
     is the estimated bolometric luminosity $L_{\rm IR}$.   
     This relation uses the
     continuous star formation model of Leitherer \et (1999) and
     a Salpeter initial mass function from 0.1 to 100 \msun. 

\label{tab:sample}  \end{table}

\clearpage

%\documentclass[manuscript]{aastex}
%\begin{document}
%\input mymac.tex

\begin{table}
\caption{Properties of Na~I Absorption}
\begin{tabular}{llllll}
\hline
 Object	& $W({\rm NaD})$	& $f_*$	     & $v_{max}$  & $C_f$ & $\Delta l$\\
		& (\AA)		& (\AA)	 	&  (\kms) &  & (${\rm h}_{70}^{-1}$ kpc)\\
\hline
\hline
IRAS00153+5454  & $4.1 \pm 0.5$ & 0.15       & -715       & 0.14 & 3.8 (7\farcs3) \\
IRAS00188-0856  & $7.5\pm0.6$	& 0.09	 	&  -751   & 0.19 & 4.6 (1\farcs9) \\
IRAS00262+4251	& $1.0\pm0.3$	& 0.33	 	& -580    & 0.11 & 6.9 (3\farcs8) \\
IRAS03158+4227  & $5.1\pm0.4$	& 0.052	 	 & -1047  & 0.25 & 6.2 (2\farcs6) \\
IRAS03521+0028	& $5.1\pm0.4$	& 0.1	      & -350      & 0.63 & 5.0 (1\farcs9) \\
IRAS08030+5243  & $7.4\pm1.0$	& 0.063	 	 & -311   & 0.26 & $\sgreat$ 8.2 (5\farcs2) \\
IRAS10494+4424	& $5.6\pm0.1$	& 0.024	 	 & -560   & 0.15 & $\sgreat$ 12.0 (7\farcs0) \\
IRAS10565+2448  & $10.4\pm0.1$	& 0.03	 	 & -620   & 0.42 & $\sgreat$ 11.4 (13\farcs4) \\
IRAS11506+1331  & $3.88\pm0.2$	& 0.10	 	 & -435   & 0.28 & 4.53 (1\farcs9)  \\
IRAS15245+1019  & $8.5 \pm 0.1$ & 0.05        & -559      & 0.67 & 4.2 (12\farcs4) \\
IRAS16090-0139	& $3.32\pm0.06$	& 0.083	 	 & -543   & 0.11 & 7.6 (3\farcs2) \\
IRAS16487+5447  & $< 0.5$      & $>0.46$      & \nodata   & \nodata & 2.71 (5\farcs7) \\
IRAS17208-0014  & $9.4\pm0.1$	& 0.06	 	 & -691   & 0.52 & $\sgreat$ 7.3 (8\farcs6) \\
IRAS18368+3549  & $9.0\pm0.3$	& 0.04	 	 & -768   & 0.25 & 10.3 (4\farcs8) \\
IRAS19297-0406	& $3.7\pm0.2$	& 0.12	 	 & -697   & 0.22 & 4.8 (2\farcs9) \\
IRAS19458+0944  & $4.0$		& \nodata   & \nodata     &  \nodata\\
IRAS20087-0308	& $9.0\pm1.5$	& 0.074	 	&   -1082 & 0.26 & 12.2 (6\farcs3) \\
IRAS23365+3604	& $4.5\pm0.4$	& 0.12	 	&   -651  & 0.16 & 5.9 (4\farcs7) \\
\hline
\end{tabular}

Table Notes --

(1) Object name.

(2) Total measured NaD absorption equivalent width (observed frame).

(3) Fraction of total NaD equivalent width from a stellar component 
 predicted from measured Mgb equivalent width and the relation
$W(\nad) = 1/3 W(Mg~I)$ described in the text.

(4)  Terminal velocity measured from the intersection of the
bluest part of the absorption line with the continuum.

(5) Fraction of continuum source covered by absorbing clouds
where $C_f \equiv 1 - I(\lambda 5890)$.

\label{tab:PROFILE} \end{table}

%\end{document}

\clearpage

%\documentclass[manuscript]{aastex}
%\begin{document}
%\input mymac.tex

\begin{table}
\caption{Fitted Na~I Absorption (Optically Thick)}
\begin{tabular}{lllllll}
\hline
 Object	         & $W_{Dop}$ & $v_{B}$	& $\Delta v_{B}$ & $\Delta v_{sys}$ & $D_B$           &$N_{NaI}$     \\				
		 & (\AA)	& (\kms)	& (\kms)    & (\kms)        &  &($10^{12}$~cm$^{-2}$)         \\				
\hline									    					   							
\hline									    					   							
IRAS00153+5454   & 3.7          & $-292\pm15$   & $452\pm40$ &  $554\pm219$   & $\sles\ 1.06\pm0.06$   & $436_{-392}^{+\inf}$ \\
IRAS00188-0856   & 4.42		& $-328\pm30$	& $520\pm20$ & 	$260\pm11$   &   $1.07\pm0.09$	& $ 310^{+\inf}_{-210}$ \\
IRAS00262+4251	 & 0.94		& $-293\pm20$	& $296\pm20$ & 	$500\pm150$  & $1.7\pm0.4$	& $1.53_{-0.25}^{+0.94}$ \\
IRAS03158+4227   & 4.43		& $-499\pm25$	& $405\pm35$ & 	$312\pm90$   & $1.21\pm0.15$	& $25_{-8}^{+12}$       \\
IRAS03521+0028	 & 5.0	        & $-77 \pm 12$	& $ 175\pm31$ & $636\pm360$  & $1.42\pm0.46$    & $1.67_{-5.4}^{270}$   \\
IRAS08030+5243   & 4.02		& $+267\pm2$	& $158\pm4$  & 	$350\pm30$   & $1.16\pm0.05$	& $29.8_{-0.85}^{+4.4}$ \\
IRAS10494+4424	 & 2.85		& $-270\pm23$   & $315\pm35$ & 	$237\pm7$   & $1.07\pm0.18$	& $174^{+\inf}_{-160}$    \\
IRAS10565+2448a  & 7.80		& $-307\pm16$	& $307\pm16$ & 	$278\pm107$  & $\sles\ 1.10$	& $\sgreat\ 160$        \\
IRAS10565+2448b  & 5.1	        & $-87\pm2$	& $188\pm10$ & 	''           & $1.33\pm0.18$	& $19.6_{-4.3}^{20.6}$  \\
IRAS11506+1331   & 3.3		& $-96\pm6$	& $242\pm10$ & 	$222\pm150$  & $1.10\pm0.07$	& $59.8_{-37.4}^{+\inf}$ \\
IRAS15245+1019   & 8.9          & $-240\pm4$    & $301\pm6$  &  $347\pm108$  & $1.24\pm0.03$  & $43.1_{-3.6}^{+5.2}$    \\
IRAS16090-0139	 & 1.31		& $-282\pm23$	& $296\pm25$ & 	$354\pm22$   & $1.96\pm0.28$	& $2.1_{-0.2}^{+4.9}$ \\
IRAS16487+5447   &  \nodata     & \nodata       & \nodata    & \nodata       &    \nodata            &   \nodata     \\
IRAS17208-0014   & 6.91		& $-409\pm3$	& $292\pm4$ & 	$254\pm6$    & $1.06\pm0.02$         & $883_{-612}^{+\inf}$  \\
IRAS18368+3549a  & 4.64		& $-420\pm35$	& $371\pm49$ & 	$364\pm134$  & $1.06\pm0.28$	& $567_{-547}^{+\inf}$  \\
IRAS18368+3549b  & 4.89		& $-143\pm4$	& $213\pm12$  & 	''   & $1.24\pm0.24$	& $22.6_{-8.1}^{+\inf}$ \\
IRAS19297-0406	 & 3.84		& $-114\pm16$	& $377\pm26$ & 	$313\pm60$  & $1.46\pm0.18$	& $11_{-1.9}^{+5.4}$  \\
IRAS19458+0944   &  \nodata     & \nodata	& \nodata    & 	 \nodata            & \nodata        & \nodata          \\
IRAS20087-0308	 & 4.3		& $-420\pm50$	& $628\pm57$ & $478\pm113$   & $1.87\pm0.36$	& $9.25_{-0.44}^{+2.95}$ \\
IRAS23365+3604	 & 1.80		& $-389\pm72$	& $308\pm85$ & $610\pm195$    & $1.26\pm0.79$	& $8.27_{-6.04}^{+\inf}$ \\
\hline
\end{tabular}

%IRAS10494+4424	 & ''		& -231 	        & $294\pm30$ & 	$212\pm15$       \\

Table Notes --

(1) Object name. Spectral model is the sum of a \nad\ doublet at the systemic velocity 
and a doublet, the dynamic component, for which the velocity is fitted.
The doublet ratio, $D \equiv W(\lambda 5890) / W(\lambda 5896)$,
of the systemic component was held at $D = 1.06$ --  equivalent
to an $\lambda 5890$ optical depth of 142.  The doublet ratio of
the dynamic component was fitted.

(2) Equivalent width of fitted Doppler-shifted component. 

(3) Fitted velocity of Doppler shifted interstellar component.

(4) FWHM of fitted Doppler shifted interstellar component.

(5) FWHM of the component at the systemic velocity.  Parentheses
indicate an assumed, rather than fitted, value.

(6) Fitted doublet ratio, $D \equiv W(5895)/W(5889)$ of Doppler shifted interstellar component.

(7) Constraints on column density from curve of growth analysis and doublet ratio from column~3.

%(8) Fraction of total fitted equivalent width in the systemic component.

\label{tab:tableFIT} \end{table}

%\end{document}

\clearpage

%\documentclass[manuscript]{aastex}
%\begin{document}
%\input mymac.tex

\begin{table}
\caption{Fitted Na~I Absorption (Optically Thin)}
\begin{tabular}{lllllll}
\hline
 Object	         & $W_{Dop}$ & $v_{B}$	& $\Delta v_{B}$ & $\Delta v_{sys}$    &$N_{NaI}$  \\
		 & (\AA)	& (\kms)	& (\kms)    & (\kms)    &  ($10^{12}$~cm$^{-2}$)  \\
\hline									    					   			
\hline									    
IRAS00153+5454   & 3.2 & $-289\pm42$   & $499\pm48$ &  $293\pm42$         & $9.94\pm0.67$     \\
IRAS00188-0856   & 4.3 & $-340\pm28$	& $567\pm36$ & 	$332\pm15$      & $ 13.1\pm0.6 $ \\
IRAS00262+4251	 & 0.5 & $-406\pm20$	& $232\pm21$ & 	$493\pm46$  	& $1.5\pm0.6$ \\
IRAS03158+4227   & 5.3 & $-469\pm11$	& $466\pm29$ & 	$340\pm83$   	& $15.67\pm0.33$ \\
IRAS03521+0028	 & 3.8 & $-94\pm15$	& $152\pm46$ & 	$530\pm200$     & $13.1\pm4.3$     \\
IRAS08030+5243   & 4.2 & $+268\pm5$	& $166\pm16$  & 	$536\pm397$   	& $13.1\pm1.1$ \\
IRAS10494+4424	 & 1.2 & $-380\pm57$ 	& $356\pm80$ & 	$318\pm15$   & $ 3.84\pm0.65$ \\
IRAS10565+2448a  & 7.7 & $-252\pm22$	& $469\pm24$ & 	$283\pm29$  	& $25.14\pm0.11$ \\
IRAS10565+2448b  & 3.9 & $-88\pm3$	& $156\pm5$ & 	''  	& $12.94\pm0.04$   \\
IRAS11506+1331   & 3.4 & $-58\pm2$	& $338\pm27$ & 	$305\pm90$  	& $10.37\pm0.02$ \\
IRAS15245+1019   & 7.4 & $-243\pm6$  & $296\pm8$  &  $307\pm20$     & $23.31\pm0.04$  \\
IRAS16090-0139	 & 0.7 & $-325\pm40$	& $281\pm37$ & 	$479\pm29$   	& $2.16\pm0.36$ \\
IRAS16487+5447   & \nodata & \nodata     & \nodata    & \nodata                &   \nodata   \\
IRAS17208-0014   & 7.4& $-359\pm15$	& $462\pm38$ & 	$253\pm13$      & $21.18\pm0.08$  \\
IRAS18368+3549a  & 5.4 & $-341\pm56$	& $469\pm66$ & 	$354\pm37$  	& $16.59\pm0.17$  \\
IRAS18368+3549b  & 3.0 & $-140\pm7$	& $168\pm19$  & ''      & $9.46\pm0.10$ \\
IRAS19297-0406	 & 3.6 & $-413\pm13$	& $370\pm22$ & 	$286\pm36$  	& $11.3\pm0.6$ \\
IRAS19458+0944   & \nodata& \nodata	& \nodata    & 	 \nodata            & \nodata   \\
IRAS20087-0308	 & 4.6 & $-421\pm50$	& $631\pm75$ & $525\pm703$   	& $14.3\pm0.8$ \\
IRAS23365+3604	 & 1.3 & $-384\pm96$	& $302\pm22$ & $660\pm179$    	& $4.3\pm1.5$ \\
\hline
\end{tabular}

%IRAS10494+4424	 & ''		& -231 	        & $294\pm30$ & 	$212\pm15$       \\

Table Notes --

(1) Object name. Spectral model is the sum of a \nad\ doublet at the systemic velocity 
and a doublet, the dynamic component, for which the velocity is fitted.
The doublet ratio was fixed at the optically thin limit.

(2) Equivalent width of fitted Doppler-shifted component. 

(3) Fitted velocity of Doppler shifted interstellar component.

(4) FWHM of fitted Doppler shifted interstellar component.

(5) FWHM of the component at the systemic velocity.  Parentheses
indicate an assumed, rather than fitted, value.

(6) Column density of Na~I.

\label{tab:Nthin} \end{table}

%\end{document}

\clearpage

%% FIGURES

% ANOTHER WAY TO INSERT FIGURES
% {\par\centering
% \resizebox*{9cm}{9cm}{\includegraphics{figures/namg.ps}}
% \par}

%FIGURE 1a
%\begin{figure}
      {\includegraphics[scale=0.7,angle=0,clip=true]{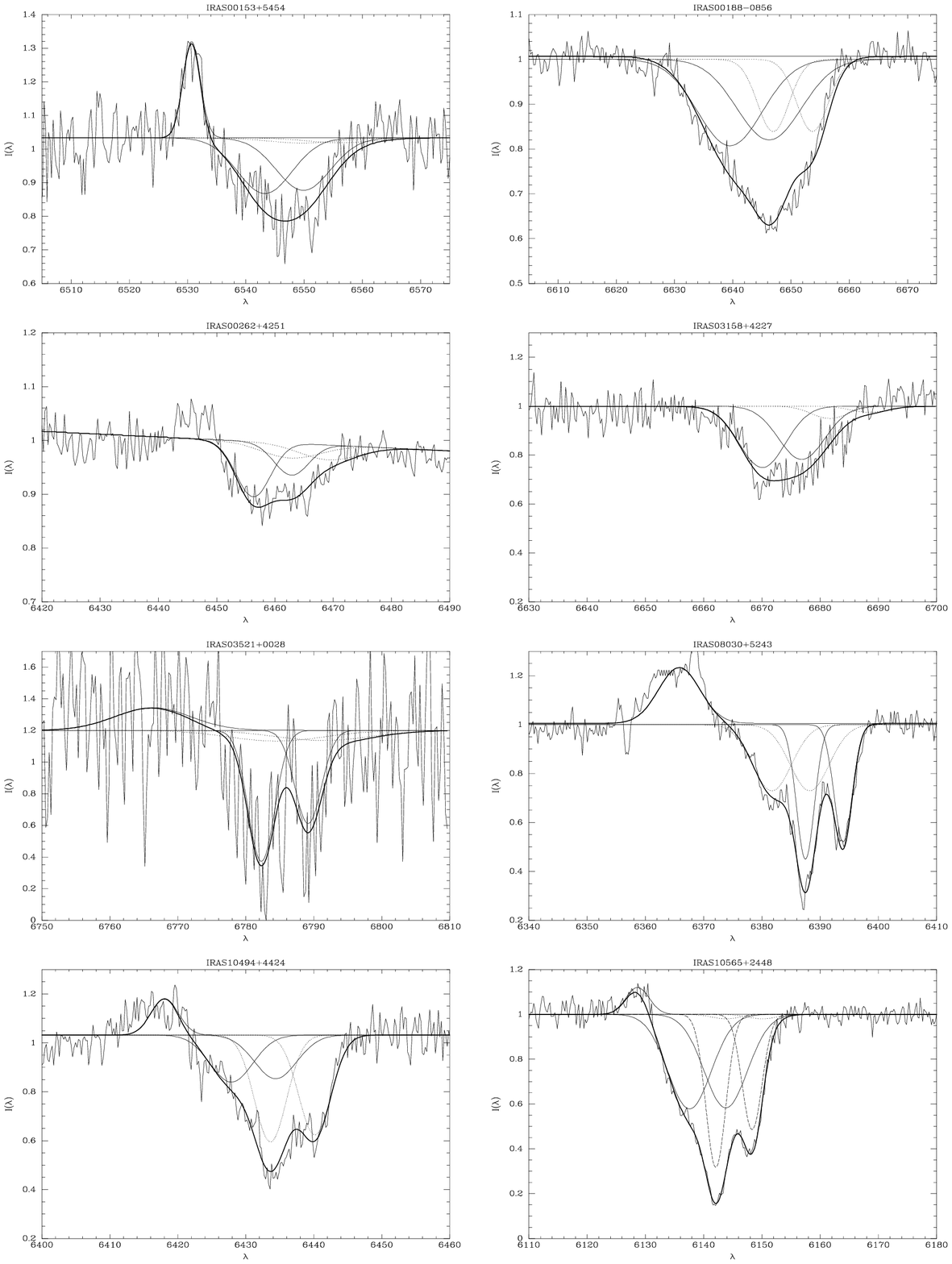}
	\hfill}
%        \caption{(a) \nad\ spectra and fitted models.}
%\label{fig:nad_pro} \end{figure}

%FIGURE 1b
\newpage

%FIGURE 2
\newpage

%FIGURE 3
\newpage

%FIGURE 4
\newpage

%FIGURE 5
\newpage

%FIGURE 6
\newpage

%FIGURE 7
\newpage

%FIGURE 8
\newpage

%FIGURE 9
\newpage
\begin{figure}
%      {\includegraphics[scale=0.7,angle=0,clip=true]{f9.eps}
%	\hfill}
        \caption{Wind strength at progressive stages of the merger.
	  The absorption lines are strongest when the merging 
	  galaxies are near perigalacticon.  They become weaker
	  around apogalacticon and again following the merger of the
	  nuclei.
	}
\label{fig:merg_seq} \end{figure}

%FIGURE 10
\newpage
\begin{figure}
      {\includegraphics[scale=0.6,angle=-90,clip=true]{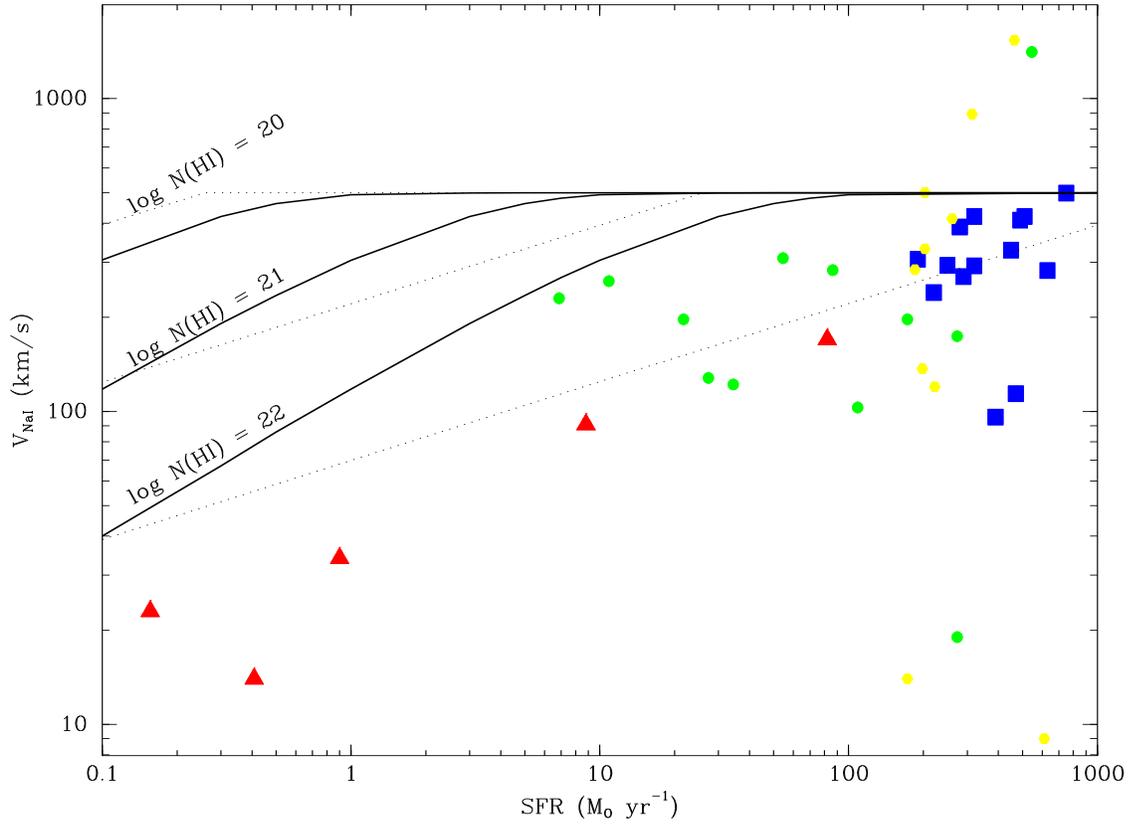}
	\hfill}
        \caption{Terminal velocity of interstellar gas clouds accelerated 
	  by a hot wind. For consistency with \x measurements, the hot wind 
	  velocity is taken as 500\kms, independent of galaxy mass; and the 
	  mass flux in the wind is assumed equal to the SFR.  Solid lines 
	  illustrate a model where the clouds are launched at a radius of 
	  200~pc. Dotted lines show the effect of increasing the size
	  of the launch region with the square root of the SFR as suggested
	  by Heckman \et (2000).  The results illustrate the critical
	  SFR below which the hot wind momentum is insufficient to
	  accelerate the clouds to the velocity of the hot wind.
	}
\label{fig:v14_sfr} \end{figure}

%FIGURE 11
\newpage
\begin{figure}
      {\includegraphics[scale=0.9,angle=-90,clip=true]{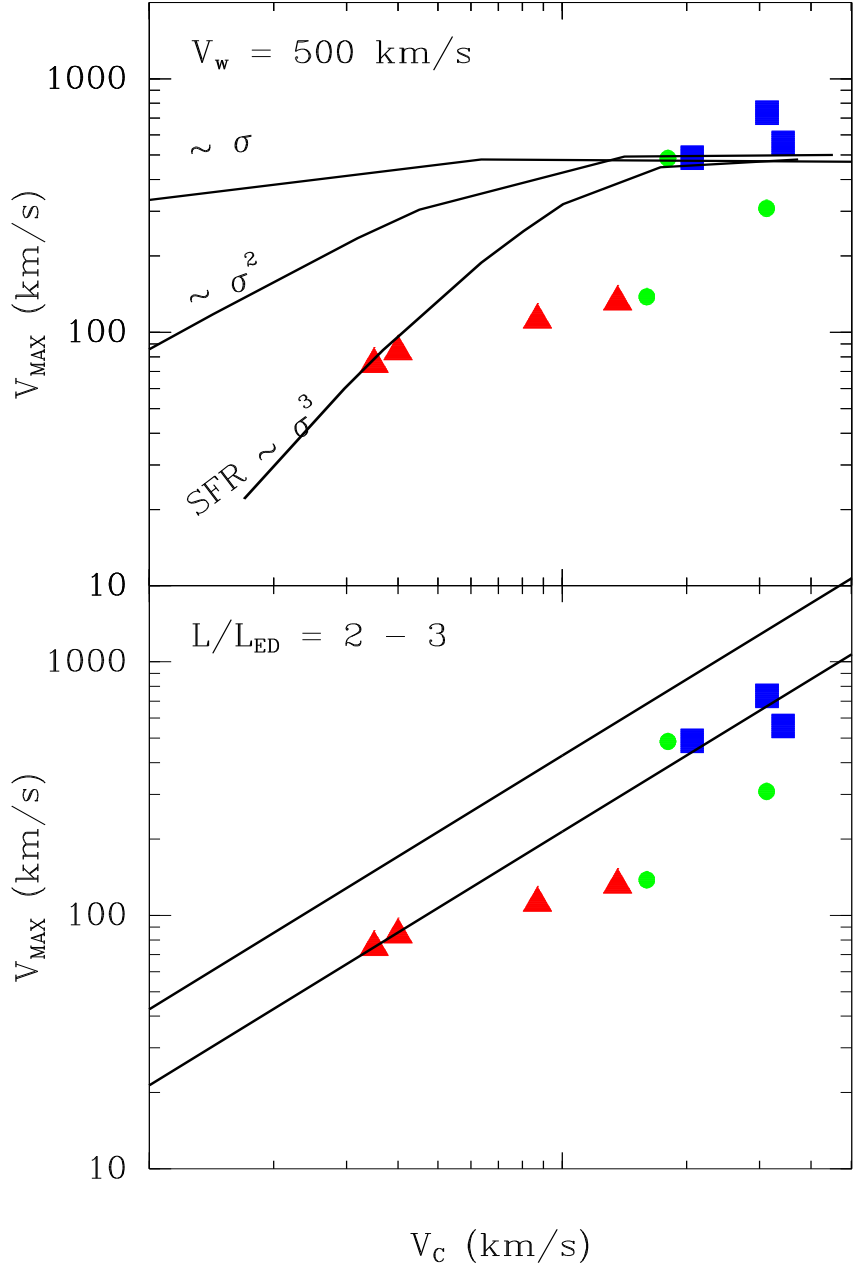}
	\hfill}
        \caption{Models for the terminal outflow velocity with respect
	  to the galactic  rotation speed.  
	  {\it (top)} The SFR increases as the 
	  cube, square, and first power of the halo velocity dispersion;
	  and momentum from the associated supernovae accelerates
	  cool clouds to the velocity of the hot wind.  The models
	  were normalized to the ULIG data such that $SFR = 500\msunyr\
	  (\sigma/210 {\rm ~km/s})^3$, 
	  $SFR = 500\msunyr (\sigma/225 {\rm ~km/s})^2$, and
	  $SFR = 500\msunyr\ (\sigma/225 {\rm ~km/s})$.
	  {\it (bottom)} The luminosity is two or three times the
	  Eddington luminosity where $L_{ED} \equiv 4 G^{-1} \sigma^4 c f_c$,
	  and the clouds, assumed to be optically thick at a radius ten times
	  the launch radius, are radiatively accelerated.  The radiative
	  acceleration model describes the data as well, and perhaps even 
	  better, than than the traditional supernova-driven wind model.
	}
\label{fig:vterm_model} \end{figure}

\end{document}